\documentclass[fleqn,usenatbib,useAMS]{mnras}

\usepackage{newtxtext,newtxmath}
\usepackage{adjustbox}
\usepackage{multirow}
\usepackage{mathrsfs}

\usepackage[T1]{fontenc}

\DeclareRobustCommand{\VAN}[3]{#2}
\let\VANthebibliography\thebibliography
\def\thebibliography{\DeclareRobustCommand{\VAN}[3]{##3}\VANthebibliography}

\usepackage{graphicx}
\usepackage{amsmath}
\usepackage{hyperref}

\title[Neutrinos and pairwise velocities]{Measuring neutrino mass and asymmetry with matter pairwise velocities}

\author[W. Z. Zhang et al.]{
Wangzheng Zhang,$^{1}$\thanks{E-mail: 1155129240@link.cuhk.edu.hk}
Ming-chung Chu,$^{1}$
Rui Hu,$^{1}$
Shihong Liao,$^{2}$
Shek Yeung$^{1}$
\\
$^{1}$ Department of Physics, the Chinese University of Hong Kong, Sha Tin, NT, Hong Kong\\
$^{2}$ Key Laboratory for Computational Astrophysics, National Astronomical Observatories, Chinese Academy of Sciences, Beijing 100101, China
}

\date{Accepted XXX. Received YYY; in original form ZZZ}

\pubyear{2023}

\begin{document}
\label{firstpage}
\pagerange{\pageref{firstpage}--\pageref{lastpage}}
\maketitle

\begin{abstract}
Neutrinos are believed to be the most abundant fermions in the Universe, but their masses are unknown, except for being non-zero but much smaller than other fermions. Cosmological relic neutrinos could also have non-zero chemical potentials (or asymmetries). Using neutrino-involved N-body simulations, we investigate the neutrino effects on the matter pairwise velocity, which itself is an interesting probe of cosmology. We find that for light-halo ($[10^{11},10^{13}]\ M_\odot$) mean pairwise velocity, in the transition range ($[4,15]\ \mathrm{Mpc}$), the effects of neutrino masses overwhelm the effects of neutrino asymmetries, while in the two-halo-group range ($[25,50]\ \mathrm{Mpc}$), for both light and heavy haloes ($[10^{13},10^{15}]\ M_\odot$), the effects of neutrino asymmetries dominate, making it possible to disentangle the two effects. We provide fitting formulae to quantify the effects of neutrino mass and asymmetry on halo-halo pairwise velocities.
\end{abstract}


%%%%%%%%%%%%%%%%%%%%%%%%%%%%%%%%%%%%%%%%%%%%%%%%%%
%%%%%%%%%%%%%%%%% BODY OF PAPER %%%%%%%%%%%%%%%%%%
\section{Introduction}

Neutrino physics carries implications beyond the Standard Model of Particle Physics. Notably, neutrino oscillation experiments unequivocally establish that at least two of the neutrino mass eigenvalues are non-zero. However, it is not known whether neutrino masses follow the normal or inverted hierarchy. From neutrino flavour oscillation experiments, assuming the normal (inverted) hierarchy, we can infer that the sum of neutrino mass eigenvalues, denoted as $M_\nu\equiv\sum_im_i$, is greater than or equal to 0.06 eV (0.10 eV) \citep{esteban2020fate}. The Karlsruhe Tritium Neutrino (KATRIN) experiment, which measures the tritium beta decay spectrum, has established an upper bound on the effective electron anti-neutrino mass, $m_{\bar\nu_e}$, at 0.8 eV (at a 90\% CL) \citep{aker2021first}, or $M_\nu\lesssim2.4\ \mathrm{eV}$. Project 8 seeks to lower the limit on $m_{\bar\nu_e}$ to 0.04 eV ($M_\nu\lesssim$ 0.12 eV) through Cyclotron Radiation Emission Spectroscopy \citep[CRES,][]{esfahani2022project}.

Beyond laboratory-based investigations, cosmology is another promising arena for studying neutrino properties, as neutrinos constitute the most abundant fermions in the universe. Relic neutrinos are radiation-like in the early universe and delay the epoch of radiation-matter equality. Consequently, they imprint discernible signatures on the cosmic microwave background (CMB). In particular, an upper bound of $M_\nu\lesssim0.26\ \mathrm{eV}$ (95\% CL) has been presented based solely on the Planck CMB data, with a refinement of $M_\nu\lesssim0.13\ \mathrm{eV}$ (95\% CL) achieved with the baryon acoustic oscillation (BAO) data included \citep{aghanim2020planck}.

In a later cosmic epoch, relic neutrinos undergo a transition to become matter-like. Nevertheless, according to \cite{shoji2010massive}, neutrinos do not cluster at scales smaller than their free-streaming length---a length inversely proportional to their masses. The high dispersion velocity inherent to neutrinos tends to wash out the formed structures, thereby imparting a notable imprint upon the large-scale structure (LSS). Combining local LSS surveys, such as the Sloan Digital Sky Survey (SDSS), Pantheon (type Ia supernovae) and Dark Energy Survey 3x2pt (DES, weak lensing) data, together with Planck CMB data mentioned above, previous works have derived  an upper bound for $M_\nu$ ranging from 0.09 to 0.115 eV (95\% CL, depending on which specific data is used and how the analysis is done) \citep{alam2021completed, valentino2021prd, palanque2020jcap}.

In the near future, several LSS surveys, such as DESI\footnote{\url{https://www.desi.lbl.gov}}, Euclid\footnote{\url{https://www.euclid-ec.org}}, LSST\footnote{\url{https://www.lsst.org}}, and SKA\footnote{\url{https://www.skao.int}}, among others, are poised to becoming invaluable tools for determining neutrino properties. While tools based on linear perturbation theory such as \texttt{CLASS} \citep{lesgourgues2011cosmic} and \texttt{CAMB} \citep{Lewis:1999bs, Howlett:2012mh} allow for qualitative understanding of neutrinos' effects on LSS, for a quantitative description of the fully non-linear evolution of LSS we have to employ N-body simulations.

One approach to incorporating neutrinos into N-body simulations involves treating them as an additional category of collisionless simulation particles, as elucidated in previous studies \citep{brandbyge2008effect, viel2010effect}. Moreover, neutrino particles possess initial velocities described by the Fermi-Dirac distribution \citep{brandbyge2008effect, viel2010effect, bird2012massive, villaescusa2014cosmology, castorina2015demnuni, inman2015precision, villaescusa2015weighing, adamek2017relativistic, emberson2017cosmological, banerjee2018reducing}. Owing to their minuscule masses and weak interaction, neutrinos exhibit substantial free-streaming behaviour, particularly at high redshifts. To accurately resolve the evolution of neutrinos in their 6D phase space, one needs to employ small time steps, which are computationally demanding. Moreover, neutrino particles are prone to multiple crossings of the simulation boundary, leading to their stochastic dispersion within the simulation box. Consequently, particle-based neutrino simulations inevitably face the shot-noise issue, as reflected in the Poisson noise on the neutrino power spectrum, scaling as $\sim 1/N_\nu$, where $N_\nu$ is the number of simulated neutrino particles. A direct approach to mitigate shot noise involves increasing $N_\nu$, as exemplified by the adoption of almost $\mathcal{O}(10^{12})$ neutrino particles in the work by \citet{emberson2017cosmological, Yu2017NatAs}. Another approach involves regular sampling of neutrino particles' velocities \citep{banerjee2018reducing}, a method implemented in the Quijote simulations \citep{villaescusa2020quijote}.

Although particle-based simulations can capture the non-linear evolution of neutrinos in the late universe, current methods for shot noise reduction impose substantial computational and memory requirements. Thus, an alternative approach employs the particle-mesh (PM) grid, where neutrinos are elaborated on the PM grid whenever long-range forces are calculated \citep{brandbyge2009grid}. Within this framework, neutrinos either evolve under a fluid Boltzmann hierarchy expansion \citep{viel2010effect, hannestad2012neutrinos, archidiacono2016efficient, banerjee2016simulating, dakin2019nuconcept, tram2019fully, inman2020simulating, chen2021cosmic} or respond linearly to a non-linear dark matter field \citep{ali2013efficient, liu2018massivenus, mccarthy2018bahamas, cartonzeng2019effects, partmann2020fast, chen2021one, hiuwing2021}. In particular, \cite{yoshikawa2020cosmological, yoshikawa2021400} directly integrate the 6D Vlasov equation of neutrinos on the grid. As there are no actual neutrino particles, grid-based neutrino-involved simulations do not have shot noise problems and also require less memory since there is no need to store the information of one additional type of particles. However, in grid-based neutrino-involved simulations, the tree force among neutrinos is neglected. Fortunately, this sacrifice is justifiable for two reasons: first, neutrinos are exceptionally light, limiting their over-density growth to around $O(1)$, and second, the scales of interest significantly exceed the mesh resolution.

Moreover, a hybrid approach, initially explored by \citet{brandbyge2010resolving}, offers an opportunity to take advantage of both grid-based and particle-based methods. In this approach, neutrinos are classified as either `fast' or `slow', each evolving under grid-based and particle-based methods, respectively \citep{banerjee2016simulating, bird2018efficient}. Another hybrid technique, also adopted in the \texttt{MillenniumTNG} \citep{hernandez2022millenniumtng} and \texttt{FLAMINGO} \citep{schaye2023FLAMINGO} simulations, solely captures deviations from a background neutrino phase space model, such as the thermal Fermi-Dirac distribution, during the evolution \citep{elbers2021optimal}.

Building upon contemporary methodologies and the capacity to execute high-resolution neutrino-involved simulations, numerous investigations have explored neutrino effects in real space. These investigations include the matter power spectrum \citep{dakin2019nuconcept, cartonzeng2019effects, elbers2021optimal}, the marked power spectrum \citep{massara2021using}, the void size function \citep{bayer2021detecting, verza2022demnuni}, the halo bias \citep{villaescusa2014cosmology, castorina2015demnuni}, the halo mass function (especially at the high-mass end, see e.g. \citealp{brandbyge2010jcap}; \citealp{castorina2015demnuni}; \citealp{adamek2017relativistic}; \citealp{liu2018massivenus}; \citealp{bayer2021detecting}), the halo merger tree \citep{liu2018massivenus, hiuwing2021}, and the cosmic neutral hydrogen (HI) distribution \citep{villaescusa2015weighing}. However, the matter velocity field, which conveys half of the phase space information was less studied. Several previous investigations have studied the impact of neutrinos on the velocity power spectrum \citep{inman2015precision, Howlett2017MNRAS, zhou2022sensitivity}. In this paper, we focus on the matter velocity field, particularly the mean peculiar pairwise velocity $v_{12}(r)$ and the mean peculiar pairwise velocity dispersion $\sigma_{12}(r)$, which have been shown to be powerful probes in constraining cosmological models.

The pairwise velocity $v_{12}(r)$ has long been used to constrain cosmological parameters such as $\sigma_8$ and $\Omega_m$ \citep{juszkiewicz2000evidence, feldman2003estimate, ma2015constraining}, the local growth rate $f\sigma_8$ \citep{Howlett2017MNRAS, dupuy2019estimation}, dark energy or modified gravity models \citep{bhattacharya2008dark, bhattacharya2011galaxy, mueller2015constraints1, bibiano2017pairwise, jaber2023arXiv231200472J}, and the kinematic Sunyaev-Zeldovich (kSZ) effects \citep{bhattacharya2007cosmological, calafut2021atacama}. Moreover, \citet{mueller2015constraints2} employed the pairwise velocity to constrain neutrino masses through the kSZ effects. The pairwise velocity dispersion $\sigma_{12}(r)$ has been used to constrain $\Omega_m$ \citep{davis1983survey, jing1998spatial, jing2002spatial, zehavi2002galaxy, hawkins20032df}, elucidate aspects of galaxy formation \citep{jing2004pairwise, li2006dependence, li2007luminosity, tinker2007luminosity, van2007towards, loveday2018galaxy}, and investigate modified gravity models \citep{hellwing2014clear}. More recently, \citet{kuruvilla2020aa} compared neutrino and baryonic effects at scale $r\lesssim20\ h^{-1}\mathrm{Mpc}$ for both $v_{12}$ and $\sigma_{12}$ in hydrodynamic simulations, while 
\citet{aviles2020jcap} studied neutrino effects at scale $r\gtrsim20\ h^{-1}\mathrm{Mpc}$ using Lagrangian perturbation theory.

In this paper, we study not only the effects of $M_\nu$ on $v_{12}(r)$ and $\sigma_{12}(r)$ but also the impact of the neutrino chemical potential ${\mu_i}$, where $i$ labels the neutrino mass eigenstates. Since the distribution function of relic neutrinos is frozen after decoupling, we define the degeneracy parameters as ${\xi_i\equiv\mu_i/T_\nu}$, where $T_\nu$ is the relic neutrino temperature. Traditionally, regardless of whether it pertains to early-universe observations, such as the CMB, or late-universe non-linear constraints on neutrinos, the analysis has assumed $\mu_i=0$ \citep{aghanim2020planck,weaklensing2023OJAp}. However, recent Planck CMB data still allow ${\xi_i}$ to assume values on the order of unity \citep{yeung2021relic}. More importantly, $\xi_i$ may alleviate the Hubble tension \citep{yeung2021relic, barenboim2017LmuLtau}. Additionally, ${\xi_i}$ play a non-negligible role in the late universe, impacting the matter power spectrum \citep{cartonzeng2019effects} and the halo merger tree \citep{hiuwing2021}, although exhibiting a high degree of degeneracy with $M_\nu$. In this study, we focus on investigating neutrino effects using a grid-based method, following the approach outlined by \citet{ali2013efficient} and \citet{cartonzeng2019effects}.

The subsequent sections of this article are organized as follows. Section \ref{sc:simulation} gives a brief review of relic neutrinos, accompanied by an elucidation of our simulation framework. Section \ref{sc:pwv} discusses the theoretical underpinnings of pairwise velocity and pairwise velocity dispersion. In Section \ref{sc:results}, we present our simulation results and quantify the neutrino effects, and Section \ref{sc:summary} summarizes our findings and discusses their implications. Appendices \ref{sc:simu_details} and \ref{sc:convergence} present the overview of our neutrino-involved simulation method and simulation convergence tests, respectively.

%%%%%%%%%%%%%%%%%%%%%%%%%%%%%%%%%%%%%%%%%%%%%%%%%%%%%%%%%%%%%%%%%%%%%%%%%%%%%%%%%%%%%
\section{Neutrino-involved simulations}\label{sc:simulation}
%%%%%%%%%%%%%%%%%%%%%%%%%%%%%%%%%%%%%%%%%%%%%%%%%%%%%%%%%%%%%%%%%%%%%%%%%%%%%%%%%%%%%

In this section, we will give a brief introduction to relic neutrinos and our neutrino-involved simulations.

%%%%%%%%%%%%%%%%%%%%%----------------subsection-----------------%%%%%%%%%%%%%%%%%%%%%
\subsection{Neutrinos}
%%%%%%%%%%%%%%%%%%%%%----------------subsection-----------------%%%%%%%%%%%%%%%%%%%%%

We assume that the relic neutrinos would be frozen in the Fermi-Dirac distribution after their decoupling from the primordial thermal bath. Consequently, the neutrino energy density is
\begin{equation}
    \rho_\nu(T_\nu) = \frac{1}{2\upi^2}\sum_{i=1}^3
    \int_0^\infty
    \left[
    \frac{E_i}{e^{(p-\mu_i)/T_\nu}+1} + \frac{E_i}{e^{(p+\mu_i)/T_\nu}+1}
    \right]
    p^2dp,
\end{equation}
where we adopt the natural units $c=\hbar=1$ and assume three mass eigenstates of neutrinos labelled by $i$. Here, $E_i=\sqrt{p^2+m_i^2}$ is the neutrino energy, and the two bracketed terms represent neutrinos and their corresponding anti-neutrinos. Furthermore, $\mu_i$ is proportional to $T_\nu$. We therefore introduce the constant factor $\xi_i\equiv \mu_i/T_\nu$, the degeneracy parameter.

Both the CMB and the LSS are sensitive only to $M_\nu$ \citep{jimenez2010can}. Previous studies, such as those by \citet{burns2023prlLe0} and \citet{escudero2023prdLe0} have established that the asymmetry between electron-type neutrinos and anti-neutrinos is severely constrained by Big Bang Nucleosynthesis (BBN), effectively yielding $|\xi_e|\lesssim0.04$. In contrast, $\xi_{\mu,\tau}\sim \mathcal{O}(1)$ remains permissible \citep{yeung2021relic}. Consequently, following \cite{cartonzeng2019effects, yeung2021relic, hiuwing2021} we set $\xi_e=0$, and given the strong mixing between $\nu_\mu$ and $\nu_\tau$, we assume $\xi_\mu=\xi_\tau$. The degeneracy parameter of neutrinos in flavour ($\xi_{e,\mu,\tau}$) and mass ($\xi_{1,2,3}$) eigenstates are related by Pontecorvo–Maki–Nakagawa–Sakata (PMNS) matrix \citep{barenboim2017prd}. This leaves us with a single independent parameter, chosen to be $\eta^2\equiv\sum \xi_i^2$. We shall henceforth refer to this parameter as the neutrino asymmetry.

Subsequently, we will perform simulations using an array of combinations of $M_\nu$ and $\eta^2$ to systematically investigate their impact on the LSS, particularly the matter pairwise velocity $v_{12}(r)$ and its corresponding dispersion $\sigma_{12}(r)$.

%%%%%%%%%%%%%%%%%%%%%----------------subsection-----------------%%%%%%%%%%%%%%%%%%%%%
\subsection{Simulations}
%%%%%%%%%%%%%%%%%%%%%----------------subsection-----------------%%%%%%%%%%%%%%%%%%%%%

Given that only $M_\nu$ is of relevance, we consider three degenerate neutrino mass eigenvalues. We have modified \texttt{Gadget-2} \citep{springel2005cosmological,cartonzeng2019effects} to incorporate neutrino effects. Our current version is almost as fast as the pure $\Lambda$CDM simulation without neutrinos. The details of the integration of neutrino effects into N-body simulations are provided in Appendix \ref{sc:simu_details}. The initial conditions for the simulations are generated using \texttt{2LPTic} \citep{crocce2006transients}, and to be consistent, we have modified the Friedmann equations used in original \texttt{2LPTic} to account for neutrino effects. The initial matter power spectra at redshift $z=99$ are derived using \texttt{CAMB} \citep{Lewis:1999bs}, with the neutrino asymmetry added \citep{yeung2021relic}. 

We have studied a total of 10 cases, which includes an even distribution of $M_\nu$ and $\eta^2$ values, coupled with the corresponding cosmological parameters as detailed in Table \ref{tb:simu_table}. To ensure consistency, for each combination of $M_\nu$ and $\eta^2$, the associated cosmological parameters have been obtained by refitting the Planck CMB and BAO data using \texttt{CosmoMC}, with neutrino asymmetry added \citep{yeung2021relic}, with the Planck 2018 plikHM\_TTTEEE likelihood. We perform two distinct simulation sets, denoted as $S_1$ and $S_2$, both with a cold dark matter (CDM) particle count of $N_p=1024^3$ and PM grid number $N_\mathrm{grid}=1024^3$, but with different box sizes of $L_{\mathrm{box}}=1000\ h^{-1}\mathrm{Mpc}$ ($S_1$) and $250\ h^{-1}\mathrm{Mpc}$ ($S_2$), respectively. Here, $h$ is the dimensionless Hubble parameter at $z=0$, i.e. $H_0 = 100 h\ {\rm km}~{\rm s}^{-1}{\rm Mpc}^{-1}$. The CDM particle mass and the gravitational softening length are around $8\times10^{10}\ h^{-1}M_\odot$ ($1\times10^{9}\ h^{-1}M_\odot$) and $24.4\ h^{-1}\mathrm{kpc}$ ($6.1\ h^{-1}\mathrm{kpc}$) for the simulation set $S_1$ ($S_2$), respectively.

The identification of dark matter haloes is performed using \texttt{ROCKSTAR} \citep{behroozi2012rockstar}, which uses adaptive hierarchical refinement of the friends-of-friends groups in six phase-space dimensions and one temporal dimension. We only consider distinct host haloes with mass defined following the virial threshold \citep{bryan1998ApJ}, i.e. $\Delta_\mathrm{vir}=102$ with respect to the critical density $\rho_\mathrm{crit}$. Moreover, we follow the definition of halo centre and velocity used in \texttt{ROCKSTAR} \citep{behroozi2012rockstar}. The halo centre is the mean value of particles in the inner subgroup of this halo, and the halo velocity is averaging particle velocity within the innermost 10\% of the halo virial radius \citep{behroozi2012rockstar}. Similar to \texttt{2LPTic}, we modify the expansion rate in \texttt{ROCKSTAR} to account for the neutrino contribution.

\begin{figure*}
    \includegraphics[width=.95\textwidth]{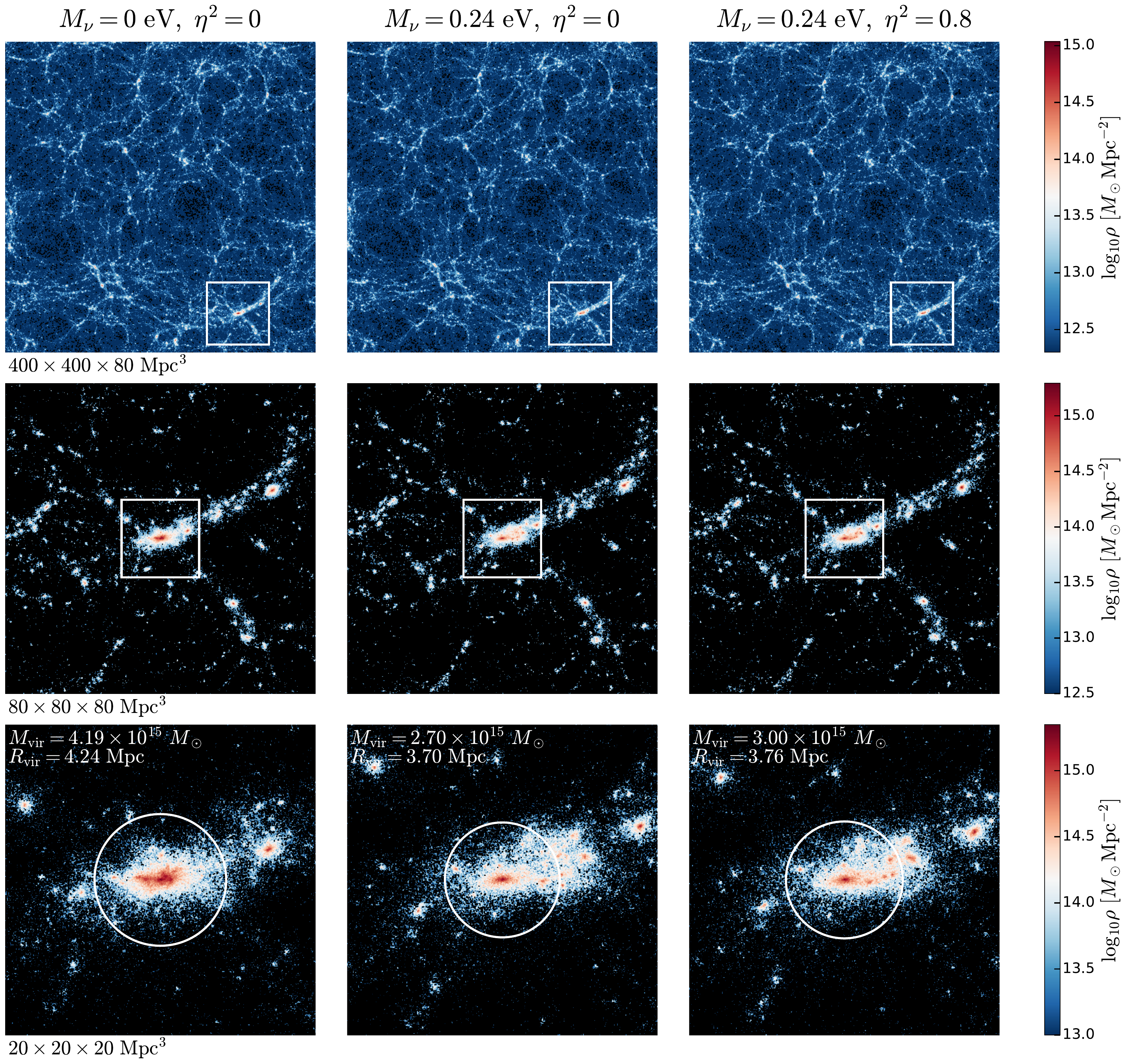}
    \caption{The 2D projected density distributions of CDM within the simulation set $S_1$ at $z=0$ are presented on various spatial scales. Each row displays a different scale, with details about the selected slices provided below. The left, middle, and right columns show the cases of $M_\nu=\eta^2=0$ (A0 in Table \ref{tb:simu_table}), $M_\nu=0.24\ \mathrm{eV}$, $\eta^2 = 0$ (A7), and $M_\nu=0.24\ \mathrm{eV}$, $\eta^2=0.8$ (A9), respectively. The second and third rows provide magnified views of the regions enclosed by white squares in the first and second rows, respectively. The last row has white circles that mark a host halo with its virial radius ($R_\mathrm{vir}$), and the corresponding $M_\mathrm{vir}$ and $R_\mathrm{vir}$ are written on the upper-left corners. The density colour bars are placed to the right of each row.}
    \label{fig:density_plot}
\end{figure*}

In Figure \ref{fig:density_plot}, we present a 2D projected density field $\rho$ from the simulation set $S_1$. The left, middle, and right columns, respectively, show the cases of $M_\nu=\eta^2=0$ (A0 in Table \ref{tb:simu_table}), $M_\nu=0.24\ \mathrm{eV}$, $\eta^2 = 0$ (A7), and $M_\nu=0.24\ \mathrm{eV}$, $\eta^2=0.8$ (A9). There are broad similarities in LSS in the different columns. However, upon closer inspection, visible differences become apparent. In particular, the finite neutrino mass serves to smooth out structures, effectively delaying structure formation. On the contrary, the introduction of a finite asymmetry results in sharper structures and an acceleration of the structure formation. This degeneracy between the effects of $M_\nu$ and $\eta^2$ on structures has been investigated in the context of the matter power spectrum \citep{cartonzeng2019effects} and the halo merger tree \citep{hiuwing2021}. In the present study, we focus on the impact of neutrinos on the matter velocity fields, representing the other half of phase space.

\begin{table*}
\large\centering
\begin{tabular}{ccccccccc}
\hline
\hline
No. &$M_\nu\ [\mathrm{eV}]$ & $\eta^2$ & $H_0\ [\mathrm{Mpc^{-1}\ km\ s^{-1}}]$ & $\Omega_{c+b}$ & $\Omega_\nu$ & $\Omega_\Lambda$ & $n_s$ & $A_s[10^{-9}]$ \\
\hline
A0 & 0    & 0   & 68.03 & 0.3070 & $\sim 10^{-5}$ & 0.6930 & 0.9661 & 2.100 \\
\hline
A1 & 0.06 & 0   & 67.73 & 0.3091 & 0.0014 & 0.6895 & 0.9667 & 2.102 \\ 
A2 & 0.06 & 0.4 & 68.77 & 0.3061 & 0.0014 & 0.6925 & 0.9728 & 2.120 \\ 
A3 & 0.06 & 0.8 & 69.88 & 0.3028 & 0.0015 & 0.6957 & 0.9789 & 2.138 \\
\hline
A4 & 0.15 & 0   & 67.14 & 0.3133 & 0.0036 & 0.6831 & 0.9680 & 2.106 \\
A5 & 0.15 & 0.4 & 68.18 & 0.3102 & 0.0037 & 0.6861 & 0.9741 & 2.123 \\
A6 & 0.15 & 0.8 & 69.26 & 0.3070 & 0.0037 & 0.6893 & 0.9802 & 2.143 \\
\hline
A7 & 0.24 & 0   & 66.55 & 0.3176 & 0.0058 & 0.6766 & 0.9691 & 2.107 \\
A8 & 0.24 & 0.4 & 67.58 & 0.3143 & 0.0059 & 0.6798 & 0.9752 & 2.127 \\
A9 & 0.24 & 0.8 & 68.62 & 0.3114 & 0.0061 & 0.6825 & 0.9814 & 2.146 \\ 
\hline
\end{tabular}
\caption{Refitted cosmological parameters employed in simulations. $M_\nu$ is the sum of the neutrino mass eigenvalues, while $\eta^2$ is the neutrino asymmetry. $H_0$ and $\Omega_i$ are the Hubble parameter and the fractional energy density attributed to the $i^{th}$ component, respectively, at redshift 0. Furthermore, $n_s$ and $A_s$ represent the spectral index and the scalar amplitude of primordial perturbations, respectively.}
\label{tb:simu_table}
\end{table*}

%%%%%%%%%%%%%%%%%%%%%%%%%%%%%%%%%%%%%%%%%%%%%%%%%%%%%%%%%%%%%%%%%%%%%%%%%%%%%%%%%%%%%
\section{Velocity fields}\label{sc:pwv}
%%%%%%%%%%%%%%%%%%%%%%%%%%%%%%%%%%%%%%%%%%%%%%%%%%%%%%%%%%%%%%%%%%%%%%%%%%%%%%%%%%%%%
In this section, we present the theoretical framework underlying the matter pairwise velocity $v_{12}(r)$, and its corresponding dispersion $\sigma_{12}(r)$.
%%%%%%%%%%%%%%%%%%%%%----------------subsection-----------------%%%%%%%%%%%%%%%%%%%%%
\subsection{Pairwise velocity}
%%%%%%%%%%%%%%%%%%%%%----------------subsection-----------------%%%%%%%%%%%%%%%%%%%%%

The mean pairwise peculiar velocity $v_{12}(r)$ is defined as 
\begin{equation}\label{eq:pwv-definition}
	v_{12}(r)\equiv\langle [\boldsymbol{v}_1(\boldsymbol{r}_1)-\boldsymbol{v}_2(\boldsymbol{r}_2)]\cdot \boldsymbol{\hat r} \rangle,
\end{equation}
where $\boldsymbol{r}\equiv\boldsymbol{r}_1-\boldsymbol{r}_2$ represents the distance vector between a pair of particles in the comoving frame, and $\boldsymbol{v}_{1,2}$ is the peculiar velocity of the respective particle. The average $\langle...\rangle$ is computed over all pairs separated by $r\equiv|\boldsymbol{r}|$, and $v_{12}(r)$ depends only on $r$ due to isotropy.

The mean pairwise velocity $v_{12}(r)$ characterizes the peculiar velocity difference between pairs of particles along their line of separation, revealing their tendency to approach each other. It is evident from Eq. (\ref{eq:pwv-definition}) that when a pair of particles move towards each other, $v_{12}<0$ (typically due to gravitational attraction). At large separation, particles have minimal correlation, so $v_{12}(r)$ tends towards zero. Conversely, for small separations, the pair of particles is either bound together or in a state of stable clustering. In this case, their mean relative velocity, denoted as $u_{12}$ in physical coordinates, approaches zero, implying \citep[\S6]{mo2010galaxy}
\begin{equation}\label{eq:v12=-Hr}
    v_{12}(r)=-H(a)ra,
\end{equation}
where $H(a)$ represents the Hubble parameter and $a$ the scale factor.

In accordance with the pair conservation equation \citep[\S6]{mo2010galaxy}, we have
\begin{equation}\label{eq:pair-conservation}
    \frac{\partial N(r,t)}{\partial t}+ 4\upi\bar na^2r^2[1+\Xi(r,a)]v_{12}(r,a)=0,
\end{equation}
where $N(r, t)$ denotes the average number of particles within a comoving distance $r$ of particle 2. The second term in the equation represents the average particle flux per unit time passing out of the spherical shell with radius $r$ centred at particle 2. Here, $\bar n(t)$ is the mean particle number density at time $t$, and $\Xi(r,a)$ is the two-point correlation function. By definition,
\begin{equation}
    N(r,t)=4\upi\bar n(t)a^3\int_0^r[1+\Xi(x,a)]x^2dx.
\end{equation}
We can simplify Eq. (\ref{eq:pair-conservation}) by utilizing the conservation of the total number of particles, $\bar n(t)a^3=\rm{constant}$, to obtain
\begin{equation}\label{eq:exact-pwv}
    \frac{v_{12}(r,a)}{Hra}=-\frac{a}{3[1+\Xi(r,a)]}\frac{\partial \bar\Xi(r,a)}{\partial a},
\end{equation}
where the averaged correlation function $\bar\Xi(r,a)$ is defined as 
\begin{equation}\label{eq:xibar}
    \bar\Xi(r,a)\equiv \frac{3}{4\upi r^3}\int d^3\boldsymbol{y} \Xi(y,a).
\end{equation}
Thus, Eq. (\ref{eq:exact-pwv}) reveals that $v_{12}(r)$ encompasses not only the information of the real space two-point correlation function $\Xi(r)$ but also its temporal evolution.

For large separation $r$, where $\Xi(r,a)\ll1$, a linear approximation can be applied. Consequently, we can expand $\Xi(r,a)$ to the first order as $\Xi(r,a)=\Xi(r,a=1)D^2(a)$, with $D(a)$ representing the linear growth solution. Eq. (\ref{eq:exact-pwv}) can then be simplified to (\citealp[\S72]{peebles2020large}; \citealp{juszkiewicz1999dynamics}),
\begin{equation}\label{eq:pwv-linear-approx}
    v_{12}(r,a)=-\frac{2}{3}Hraf(a)\bar{\bar\Xi}(r,a),
\end{equation}
where $\bar{\bar{\Xi}}(r,a)\equiv \bar\Xi(r,a)/[1+\Xi(r,a)]$, and the linear growth rate $f(a)$ is defined as $\mathrm{d}\ln D/\mathrm{d}\ln a$. A reasonable approximation for $f(a)$ is $\Omega_m^{0.55}(a)$, with $\Omega_m(a)$ representing the fractional energy density of the total matter \citep{linder2005Omega0p55}.

%%%%%%%%%%%%%%%%%%%%%----------------subsection-----------------%%%%%%%%%%%%%%%%%%%%%
\subsection{Pairwise velocity dispersion}\label{sc:intro-sigma12}
%%%%%%%%%%%%%%%%%%%%%----------------subsection-----------------%%%%%%%%%%%%%%%%%%%%%
Similarly, the mean pairwise peculiar velocity dispersion is defined as
\begin{equation}\label{eq:sigma12}
    \sigma_{12}(r)
    \equiv
    \langle 
    \{
    [\boldsymbol{v}_1(\boldsymbol{r}_1)-\boldsymbol{v}_2(\boldsymbol{r}_2)]\cdot \boldsymbol{\hat r}
    \}^2 
    \rangle ^{1/2}.
\end{equation}

By integrating over momenta of the second Bogoliubov–Born–Green–Kirkwood–Yvon (BBGKY) equation and following a series of simplifications, we arrive at \citep[\S72]{peebles2020large}
\begin{equation}\label{eq:sigma12-eqution}
    \begin{aligned}
   	\frac{\partial}{\partial t}(1+\Xi)\langle \tilde{v}_{12}^i \rangle 
   		&+H(1+\Xi)\langle \tilde{v}_{12}^i \rangle\\
   		&+\frac{1}{a}\sum_j\frac{\partial}{\partial r^j}(1+\Xi)\langle \tilde{v}_{12}^i \tilde{v}_{12}^j \rangle\\
   		&+\frac{2Gm}{a^2}\frac{r^i}{r^3}(1+\Xi)
   		+2Gm\bar n a\frac{r^i}{r^3}\int_0^r\Xi d^3\boldsymbol{r}\\
   		&+2Gm\bar n a\int\zeta(1,2,3)\frac{r_{13}^i}{r_{13}^3}d^3\boldsymbol{r}_3=0,\\
    \end{aligned}
\end{equation}
where $\boldsymbol{r}_{13}\equiv\boldsymbol{r}_1-\boldsymbol{r}_3$ and $\boldsymbol{\tilde{v}}_{12} \equiv \boldsymbol{v}_1-\boldsymbol{v}_2$. $\zeta(1,2,3)$ is the three-point correlation function and $m$ the particle mass. This equation describes how velocity dispersion and gravity affect the evolution of $\Xi$. The third to last and final two terms are the gravitation from the particle pair and all neighboring particles, respectively.

The mean pairwise peculiar velocity dispersion $\sigma_{12}$ bears a direct relation to the term $\langle \tilde{v}_{12}^i\tilde{v}_{12}^j \rangle$ within Eq. (\ref{eq:sigma12-eqution}). This quantity can be expressed as 
\begin{equation}\label{eq:dispersion-tensor}
    \langle \tilde{v}_{12}^i\tilde{v}_{12}^j \rangle
    =
    \left(
    \frac{2}{3}\langle v_1^2 \rangle +\Sigma 
    \right)
    \delta^{ij}+(\Pi-\Sigma)\frac{r^ir^j}{r^2},
\end{equation}
where the $2\langle v_1^2 \rangle/3$ is due to the uncorrelated motion between the particles, while $\Pi$ and $\Sigma$ characterize the effects of correlated motions along the components of dispersion parallel and perpendicular to the separation vector $\boldsymbol{r}$, respectively. Thus, when projecting Eq. (\ref{eq:dispersion-tensor}) on the separation line, we obtain $\sigma_{12}^2=\frac{2}{3}\langle v_1^2 \rangle+\Pi$.

At small separation scales, the system assumes a statistically static state, allowing for an estimation of $\sigma_{12}(r)$ via the Cosmic Virial Theorem (CVT) derived from Eq. (\ref{eq:sigma12-eqution}) under the condition $\zeta\gg\Xi\gg 1$ (\citealp[\S6]{mo2010galaxy}; \citealp[\S75]{peebles2020large}):
\begin{equation}\label{eq:CVT}
    \sigma_{12}^2(r)
    =
    \frac{3\Omega_m(a) H^2(a)}{4\upi\Xi(r,a)}\int_r^\infty\frac{dr'}{r'}\int d^3\boldsymbol{z}\frac{\boldsymbol{r}'\cdot \boldsymbol{z}}{z^3}\zeta(r',z,|\boldsymbol{r}'-\boldsymbol{z}|,a).
\end{equation}
Therefore, the CVT serves as a pivotal link between the mean relative velocity squared (representing kinetic energy) and the mean gravitational potential energy originating from all neighboring pairs.

When $r$ is large, 
\begin{equation}\label{eq:CEE}
    \sigma_{12}^2(r)\approx2\langle v_1^2 \rangle/3,    
\end{equation}
where the factor 2 arises from the pairing, and $1/3$ signifies that only the direction along the line of separation is considered. In parallel, $\langle v_1^2 \rangle$ can be estimated using the Cosmic Energy Equation (CEE; \citealp[\S6]{mo2010galaxy}; \citealp[\S74]{peebles2020large}; \citealp{mo1997analytical}), which stems from the integration of the first BBGKY equation over the momentum space, 
\begin{equation}
    \frac{d}{da}a^2\langle v_1^2 \rangle=4\upi G\bar\rho a^3\frac{\partial \mathcal{I}_2(a)}{\partial \ln a},
\end{equation}
where $\mathcal{I}_2\equiv\int_0^\infty dy\Xi(y,a)y$. Integrating the above equation yields
\begin{equation}\label{eq:CEE_v2}
    \langle v_1^2 \rangle = \frac{3}{2}\Omega_m(a)H^2(a)a^2\mathcal{I}_2(a)\left[ 1-\frac{1}{a\mathcal{I}_2(a)}\int_0^a da' \mathcal{I}_2(a') \right].
\end{equation}
Notably, the CEE establishes a connection between the mean-squared peculiar velocity (representing kinetic energy) and its associated mean gravitational potential energy. In essence, the CEE embodies the conservation of Newtonian energy within the expanding universe.

As shown in \cite{cartonzeng2019effects, hiuwing2021}, the neutrino mass and asymmetry affect the large-scale matter distribution and structure growth rates, based on the theoretical framework outlined above, we expect that they will also affect $v_{12}$ and $\sigma_{12}$. In the following section, we will explicitly test this theoretical framework in the neutrino case by comparing the theoretical calculations with the simulation results.

%%%%%%%%%%%%%%%%%%%%%%%%%%%%%%%%%%%%%%%%%%%%%%%%%%%%%%%%%%%%%%%%%%%%%%%%%%%%%%%%%%%%%
\section{Results}\label{sc:results}
%%%%%%%%%%%%%%%%%%%%%%%%%%%%%%%%%%%%%%%%%%%%%%%%%%%%%%%%%%%%%%%%%%%%%%%%%%%%%%%%%%%%%

In this section, we will discuss our results of $v_{12}(r)$ and $\sigma_{12}(r)$, both at $z = 0$, obtained with the data sets $S_1$ ($L_{\mathrm{box}}=1000\ h^{-1} \mathrm{Mpc}$) and $S_2$ ($L_{\mathrm{box}}=250\ h^{-1}\mathrm{Mpc}$), as described in Section \ref{sc:simulation}. There are three effects of relic neutrinos: the modified expansion history, the neutrino free-streaming effects, and the refitted cosmological parameters. According to \citet{cartonzeng2019effects}, the modified expansion history has a negligible, sub-percentage impact on the matter power spectrum. On the other hand, the effects from neutrino free-streaming and the refitted cosmological parameters are more important.

In Appendix \ref{sc:nu_cosmic_variance}, we investigate the impact of cosmic variance (i.e. different random seeds to generate initial conditions) on the particle-particle (halo-halo) pairwise velocity, $v_\mathrm{pp}$ ($v_\mathrm{hh}$), and its associated dispersion, $\sigma_\mathrm{pp}$ ($\sigma_\mathrm{hh}$). We find that even though there are fluctuations in the magnitudes of $v_\mathrm{pp/hh}$ and $\sigma_\mathrm{pp/hh}$, the effect of neutrinos remains robust and unaffected by cosmic variance. Our results presented below are based on simulations using the same random seed.

%%%%%%%%%%%%%%%%%%%%%----------------subsection-----------------%%%%%%%%%%%%%%%%%%%%%
\subsection{Particle-particle case}\label{sc:particle-particle-pwv}
%%%%%%%%%%%%%%%%%%%%%----------------subsection-----------------%%%%%%%%%%%%%%%%%%%%%

As demonstrated in Appendix \ref{sc:pp_pert_part}, it is accurate enough to randomly select 1\% of the total CDM particles to calculate the particle-particle pairwise velocity ($v_\mathrm{pp}$) and dispersion ($\sigma_\mathrm{pp}$), with uncertainties staying within 0.5\% when $r\gtrsim4\ \mathrm{Mpc}$. Additionally, from Appendix \ref{sc:pp_boxsize}, the neutrino effects based on simulation sets $S_1$ and $S_2$ are comparable. Therefore, we will only present $v_\mathrm{pp}$ and $\sigma_\mathrm{pp}$ using 1\% of randomly chosen CDM particles from $S_1$.

\begin{figure*}
    \includegraphics[width=.95\textwidth]{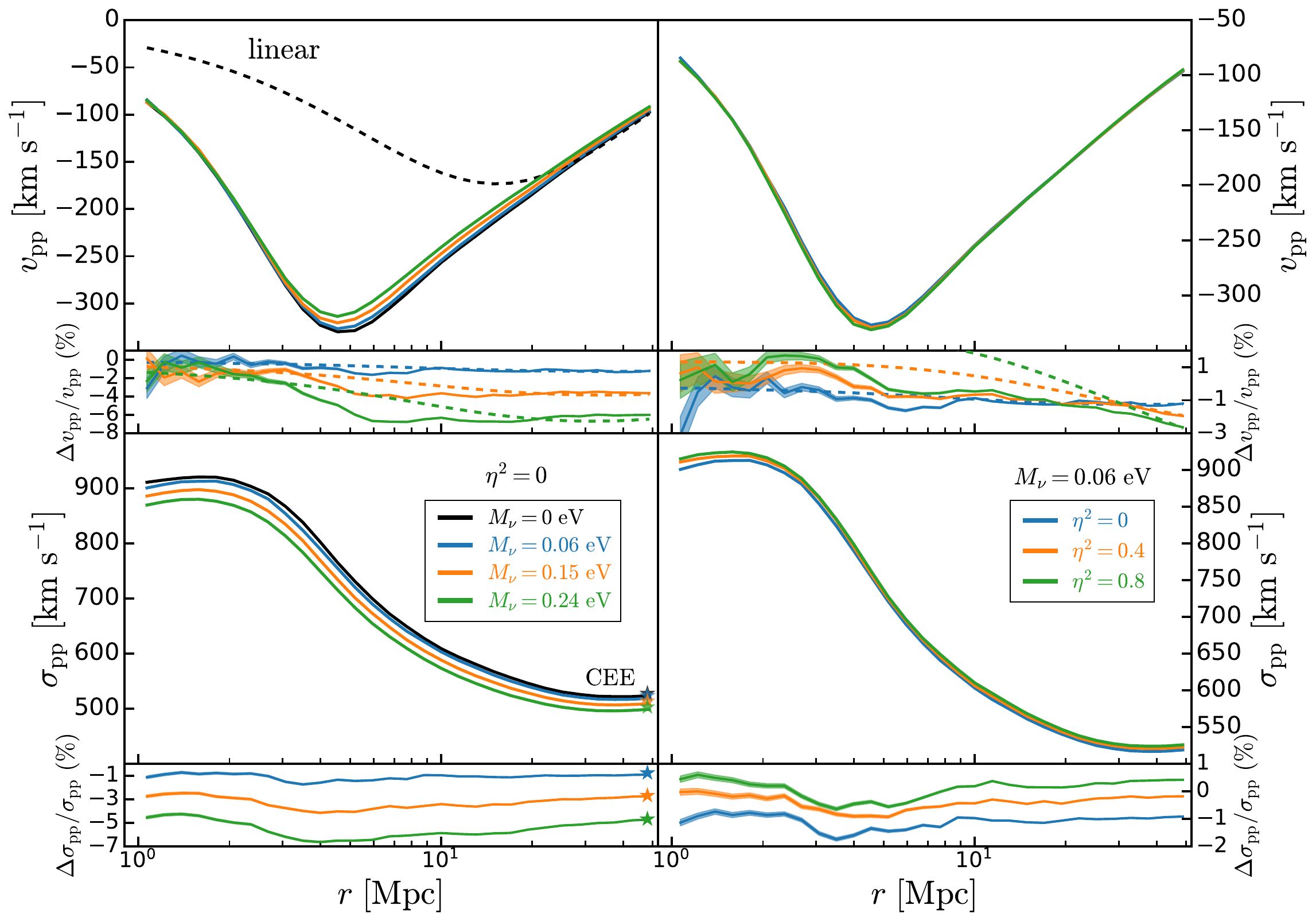}
    \caption{Particle-particle mean pairwise peculiar velocity ($v_{\mathrm{pp}}$, upper panels) and dispersion ($\sigma_\mathrm{pp}$, lower panels), respectively, for $\eta^2=0$ and different $M_\nu$  (left panels), as well as $M_\nu=0.06\ \mathrm{eV}$ and different $\eta^2$ (right panels). The subpanels below each graph show the percentage fractional deviations between the coloured cases and the reference case with massless neutrinos (A0 in Table \ref{tb:simu_table}). The upper panels show the linear approximation in dashed lines, Eq. (\ref{eq:pwv-linear-approx}), and the coloured stars in the lower left plot show the results from the CEE, as in Eq. (\ref{eq:CEE}).}\label{fig:pp-pwv}
\end{figure*}

Figure \ref{fig:pp-pwv} displays $v_\mathrm{pp}$ ($\sigma_\mathrm{pp}$) in the top (bottom) row. The subpanels show the percentage fractional deviations from the reference scenario with massless neutrinos (A0 in Table \ref{tb:simu_table}). The first column keeps $\eta^2=0$ while changing $M_\nu$, and the second column varies $\eta^2$ while keeping $M_\nu=0.06\ \mathrm{eV}$.

The dashed lines in the top row of the subpanel are derived from the linear approximation of $v_\mathrm{pp}$ given by Eq. (\ref{eq:pwv-linear-approx}), where $\Xi$ is obtained by the inverse Fourier transformation (IFT) of the linear matter power spectrum at $z=0$ (generated from \texttt{CAMB}) with the same cosmological parameters listed in Table \ref{tb:simu_table}. Our analysis shows that the simulation result (the solid black line) is in close agreement with the linear approximation (the dashed black line) for large $r$ in the upper left subpanel. Additionally, in the lower subpanels of the top row, the neutrino effects from simulations (solid coloured lines) agree with the corresponding linear approximations (dashed coloured lines) at large $r$. The linear theory consistently underestimates the behaviour of $v_\mathrm{pp}$ and fails to provide meaningful insight into the non-linear regime, which encompasses nearly all of the $r$ values depicted in Figure \ref{fig:pp-pwv}.

The coloured stars in the lower left panel of Figure \ref{fig:pp-pwv} represent the results obtained through the CEE, as expressed in Eq. (\ref{eq:CEE}). Here, $\Xi$ in $\mathcal{I}_2$ (i.e. in Eq. (\ref{eq:CEE_v2})) is the IFT of the non-linear halofit matter power spectrum \citep{bird2012massive} (chopping $k$ lower than $2\upi/L_\mathrm{box}$ to account for the finite box size effect). Due to the lack of a halofit model that takes into account the finite asymmetry $\eta^2$, we only include the cases with $\eta^2$ fixed at zero. The lower left panel shows that the values of $\sigma_\mathrm{pp}$ calculated from CEE are in close agreement with our simulation results. In particular, the effects of neutrinos, as shown in the lower subpanel, are in good agreement with the predictions of the CEE.

The upper panels of Figure \ref{fig:pp-pwv} show that the minimum of $v_\mathrm{pp}$ is at $r \sim 4.5\ \mathrm{Mpc}$, which is the transition scale between the one-halo and two-halo regions. As $r$ increases or decreases, the pairwise velocity tends to zero, which is consistent with our theoretical predictions. At large separations, CDM particles move randomly with little correlation, whereas at smaller separations, they tend to form stable clusters.

In the first column of Figure \ref{fig:pp-pwv}, it is clear that as $M_\nu$ increases, both the magnitudes of $v_\mathrm{pp}$ and $\sigma_\mathrm{pp}$ decrease. This is in line with the idea that neutrinos tend to smooth out LSS. Additionally, when comparing the left and right columns of the figure, it is evident that $M_\nu$ and $\eta^2$ have opposite effects on $v_\mathrm{pp}$ and $\sigma_\mathrm{pp}$, except for $v_\mathrm{pp}$ at large $r$. In this case, a larger $\eta^2$ reduces $v_\mathrm{pp}$, same as increasing neutrino masses. In summary, $\eta^2$ increases the magnitude of $v_\mathrm{pp}$ in the region between the one-halo and two-halo regimes and smooths out $v_\mathrm{pp}$ in the two-halo region. However, interestingly, the effects of $\eta^2$ on $\sigma_\mathrm{pp}$ remain in the opposite direction to $M_\nu$ throughout the range we show.

To conduct a quantitative exploration analysis of these phenomena, we introduce the ratio of $v_\mathrm{pp}$, denoted as $R^v_{\rm{pp}}$,  
\begin{equation}
    R^v_{\rm{pp}}(M_\nu,\eta^2,r)
        \equiv
    \frac{v_{\mathrm{pp}}(M_\nu,\eta^2,r)}{v_{\mathrm{pp}}(0,0,r)},
\end{equation}
where $v_{\mathrm{pp}}(0,0,r)$ is from the fiducial massless scenario (A0 in Table \ref{tb:simu_table}).

\begin{figure*}
    \includegraphics[width=0.95\textwidth]{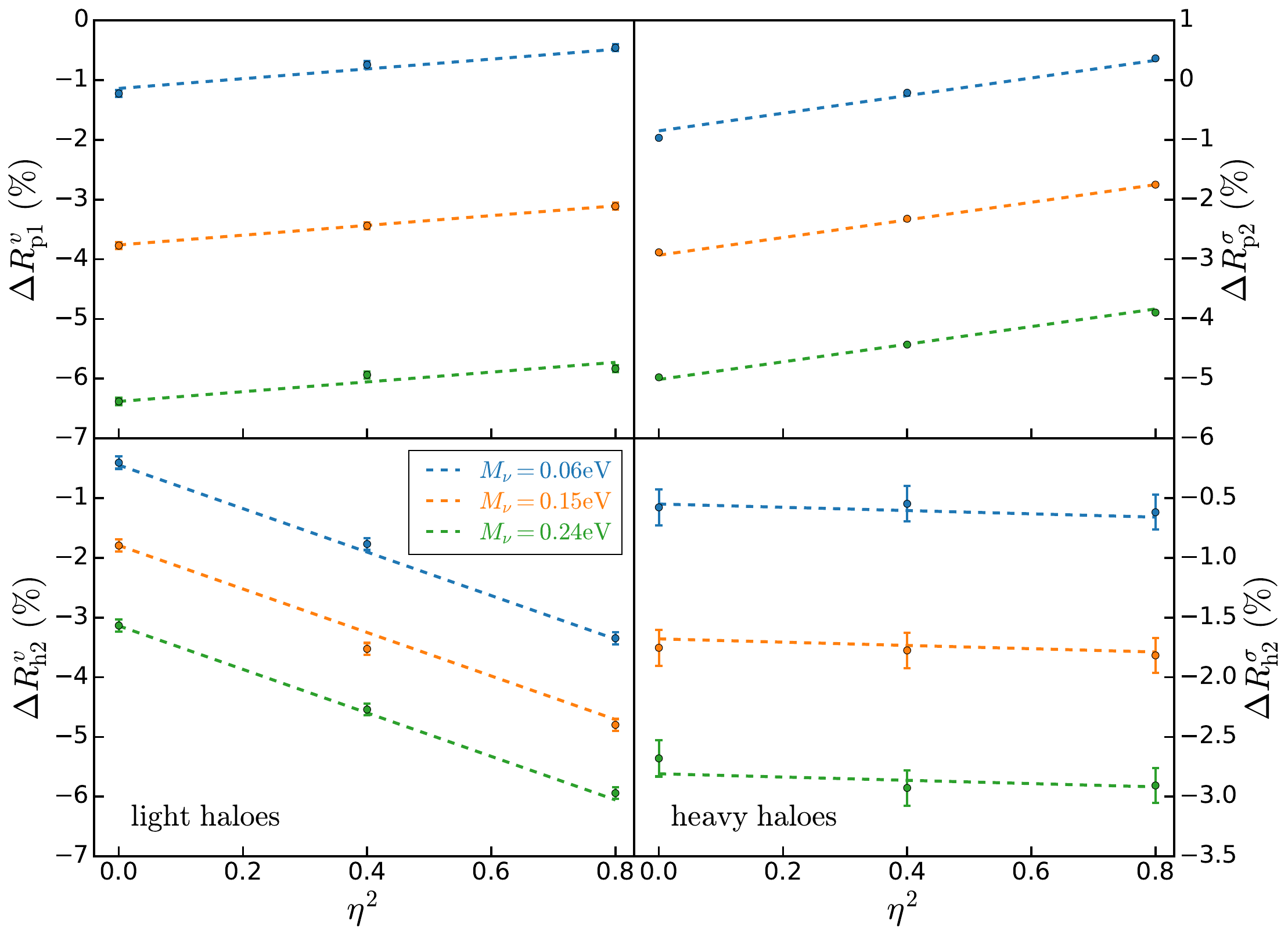}
    \caption{$\Delta R^v_{\mathrm{p}1}$ (upper left panel) and $\Delta R^\sigma_{\mathrm{p}2}$ (upper right panel) as a function of $\eta^2$ for different $M_\nu$. The lower left and right panels show $\Delta R^v_{\mathrm{h}2}$ (from light haloes) and $\Delta R^\sigma_{\mathrm{h}2}$ (from heavy haloes), respectively. The data points acquired from simulations, each with its own error bars, are fitted to the two-variable linear regression model (dashed lines), and the different colours show different $M_\nu$.}\label{fig:pp-hh-v12-d12-fitting-results}
\end{figure*}

Subsequently, we can define the averaged value of $R_\mathrm{pp}^v(M_\nu,\eta^2,r)$ within a certain $r$ interval $I_j$ as $\Delta R^v_{\mathrm{p}j}(M_\nu,\eta^2)$,
\begin{equation}
    \Delta R^v_{\mathrm{p}j}(M_\nu,\eta^2) 
    \equiv 
    \left.\overline{ R^v_{\mathrm{pp}}(M_\nu,\eta^2,r) }\right\vert_{r\in I_j} - 1,
\end{equation}
where $j$ denotes different $r$ intervals.

The effects of neutrinos are seen mainly at $r$ beyond the one-halo region. Also, as Appendix \ref{sc:pp_boxsize} shows, the effects of neutrinos do not converge within the one-halo range under the current simulation resolution. Therefore, we consider two different $r$ intervals: $I_1=[4,15]\ \mathrm{Mpc}$ ($j=1$) and $I_2=[25,50]\ \mathrm{Mpc}$ ($j=2$).

Furthermore, we apply the two-variable linear regression on $\Delta R^v_{\mathrm{p}j}(M_\nu,\eta)$,
\begin{equation}\label{eq:pp-pwv-fitting-formula}
    \Delta R^v_{\mathrm{p}j}(M_\nu,\eta^2)
    =
    C^{v}_{m\mathrm{p}j} M_\nu'+C^v_{\eta\mathrm{p}j} \eta^2,
\end{equation}
where $C^{v}_{m\mathrm{p}j}$ and $C^v_{\eta\mathrm{p}j}$ are two fitting parameters, and $M_\nu'\equiv M_\nu/0.06\ \mathrm{eV}$ denotes the dimensionless neutrino mass.

By fitting our simulation data to this linear regression model (Eq. (\ref{eq:pp-pwv-fitting-formula})), we obtain the best-fitting values for $r\in [4,15]\ \mathrm{Mpc}$: $C^{v}_{m\mathrm{p}1}(\%)=-1.596\pm0.010$ and $C^v_{\eta\mathrm{p}1}(\%)=1.157\pm0.053$. In this region, when $M_\nu'\sim O(1)$ and $\eta^2\sim O(1)$, the effect of neutrino mass slightly overwhelms that of the asymmetry, and they act in opposite directions. Conversely, in the two-halo region, we find that the effect of neutrino mass remains relatively unchanged, while that of the asymmetry changes sign ($C^{v}_{m\mathrm{p}2}(\%)=-1.530\pm0.003$ and $C^v_{\eta\mathrm{p}2}(\%)=-0.906\pm0.014$). The fitting results are summarized in Table \ref{tb:fitting-results-pp}.

Similarly, in the case of $\sigma_{\mathrm{pp}}$, we can define
\begin{equation}
    R^\sigma_{\rm{pp}}(M_\nu,\eta,r)
        \equiv
    \frac{\sigma_{\mathrm{pp}}(M_\nu,\eta,r)}{\sigma_{\mathrm{pp}}(0,0,r)}.
\end{equation}
Subsequently, we employ the same linear regression approach for $v_\mathrm{pp}$ to $\Delta R^\sigma_{\mathrm{p}j}\equiv\left.\overline{ R^\sigma_{\mathrm{pp}}(M_\nu,\eta,r) }\right\vert_{r\in I_j} - 1$,
\begin{equation}\label{eq:pp-d12-fitting-formula}
    \Delta R^\sigma_{\mathrm{p}j}(M_\nu,\eta)
    =
    C^{\sigma}_{m\mathrm{p}j} M_\nu'+C^\sigma_{\eta\mathrm{p}j} \eta^2.
\end{equation}

The results of the fitting are also shown in Table \ref{tb:fitting-results-pp}. In the transition region $[4,15]\ \mathrm{Mpc}$, when $M_\nu'\sim O(1)$ and $\eta^2\sim O(1)$, the neutrino mass and asymmetry have similar effects on $\sigma_{\mathrm{pp}}$ ($C^\sigma_{m\mathrm{p}1}(\%)=-1.538\pm0.006$ and $C^\sigma_{\eta\mathrm{p}1}(\%)=1.569\pm0.032$), but with opposite signs. In the two-halo region $[25,50]\ \mathrm{Mpc}$, the effect of $\eta^2$ is slightly more prominent than that of $M_\nu$ ($C^\sigma_{m\mathrm{p}2}(\%)=-1.254\pm0.001$ and $C^\sigma_{\eta\mathrm{p}2}(\%)=1.775\pm0.004$). Additionally, the effects of $\eta^2$ on $\sigma_\mathrm{pp}$ remain consistent in direction for both the transition and two-halo regions, unlike the case for $v_\mathrm{pp}$. The upper left and right panels of Figure \ref{fig:pp-hh-v12-d12-fitting-results} show the fitting results for $\Delta R^v_{\mathrm{p}1}$ and $\Delta R^\sigma_{\mathrm{p}2}$.

\begin{table}
\centering
  \begin{tabular}{cccc}
    \hline
    \hline
    range [Mpc] & quantity 
    & $C^{v/\sigma}_{m\mathrm{p}j}$ (\%)& $C^{v/\sigma}_{\eta\mathrm{p}j}$ (\%)\\
    \hline
    \multirow{2}{*}{[4,15]} 
    & \multicolumn{1}{c}{$v_{\mathrm{pp}}$}      & \multicolumn{1}{c}{$-1.596\pm0.010$} & \multicolumn{1}{c}{$1.157\pm0.053$} \\
    & \multicolumn{1}{c}{$\sigma_{\mathrm{pp}}$} & \multicolumn{1}{c}{$-1.538\pm0.006$} & \multicolumn{1}{c}{$1.569\pm0.032$} \\
    \hline
    \multirow{2}{*}{[25,50]} 
    & \multicolumn{1}{c}{$v_{\mathrm{pp}}$}      & \multicolumn{1}{c}{$-1.530\pm0.003$} & \multicolumn{1}{c}{$-0.906\pm0.014$} \\
    & \multicolumn{1}{c}{$\sigma_{\mathrm{pp}}$} & \multicolumn{1}{c}{$-1.254\pm0.001$} & \multicolumn{1}{c}{$1.775\pm0.004$}  \\    
    \hline
  \end{tabular}
\caption{Linear regression fitting results of $\Delta R^v_{\mathrm{p}j}$ and $\Delta R^\sigma_{\mathrm{p}j}$.}
\label{tb:fitting-results-pp}
\end{table}

%%%%%%%%%%%%%%%%%%%%%----------------subsection-----------------%%%%%%%%%%%%%%%%%%%%%
\subsection{Halo-halo case}\label{sc:halo-halo-pwv}
%%%%%%%%%%%%%%%%%%%%%----------------subsection-----------------%%%%%%%%%%%%%%%%%%%%%

The simulation sets $S_1$ and $S_2$ have different halo mass distributions. We choose host haloes from each set that fall within two particular virial mass intervals: $[10^{11},10^{13}]\ M_\odot$ (referred to as "light haloes") from $S_2$, and $[10^{13},10^{15}]\ M_\odot$ (referred to as "heavy haloes") from $S_1$. This selection yields a population of around $250,000$ haloes for each case, with each halo composing of at least 200 CDM particles.

\begin{figure*}
    \includegraphics[width=.95\textwidth]{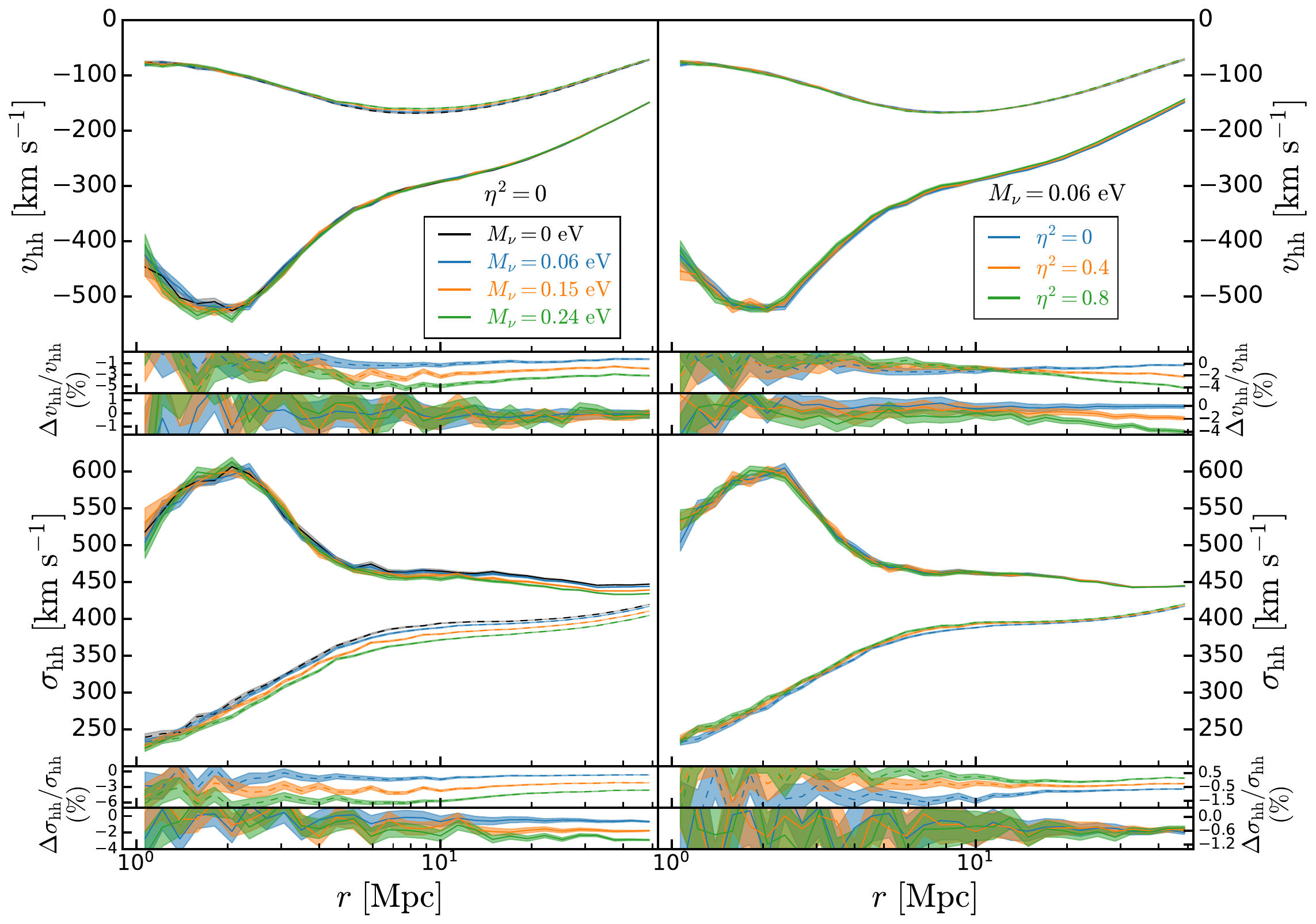}
    \caption{Halo-halo pairwise velocity $v_{\mathrm{hh}}$ (upper panels) and dispersion $\sigma_{\mathrm{hh}}$ (lower panels), respectively, for heavy haloes ($S_1$, solid lines) and light haloes ($S_2$, dashed lines) $\eta^2=0$ and different $M_\nu$ (left panels), as well as $M_\nu=0.06\ \rm{eV}$ and different $\eta^2$ (right panels). The first (light haloes) and second (heavy haloes) subpanels below each graph show the percentage fractional deviations between the massive and the corresponding massless neutrino cases.}
    \label{fig:hh-pwv}
\end{figure*}

The upper panels of Figure \ref{fig:hh-pwv} show that $v_{\mathrm{hh}}$ for light haloes (dashed lines, mass range $[10^{11},10^{13}]\ M_\odot$) is consistently less negative than for heavy haloes (solid lines, mass range $[10^{13},10^{15}]\ M_\odot$). This may be due to two reasons: firstly, the stronger gravitational interaction among heavier haloes, secondly, halo bias, i.e. heavier haloes tend to be located in denser regions, resulting in larger magnitudes of pairwise velocities. The same pattern is visible in the lower panels of Figure \ref{fig:hh-pwv}, where $\sigma_{\mathrm{hh}}$ for light haloes is smaller than for heavy haloes.

Furthermore, it is evident from the upper left panel of Figure \ref{fig:hh-pwv} that, for two different ranges of halo mass, the minimum points of $v_{\mathrm{hh}}(r)$, denoted as $(r_{\mathrm{min}}, v_{\mathrm{min}})$, show considerable differences. For light haloes, $r_{\mathrm{min}}\sim8\ \mathrm{Mpc}$, which is the transition radius between the regions of one halo group and two halo groups. This is similar to the behaviour observed for $v_\mathrm{pp}$, although the transition radius is larger for haloes than for particles. In the vicinity of $r_{\mathrm{min}}$ for light haloes, an increase in neutrino mass leads to an increase in $v_{\mathrm{min}}$, while $v_{\mathrm{min}}$ is much less sensitive to $\eta^2$. These trends are also visible in $\sigma_{\mathrm{hh}}$ (dashed lines in the lower panels).

Additionally, from the subpanels of Figure \ref{fig:hh-pwv}, it is evident that the pairwise velocity of heavy and light haloes is more sensitive to $\eta^2$ for $r\in[30,50]\ \mathrm{Mpc}$ (the two-halo-group region). However, this sensitivity is reduced for small $r$ centered around light haloes $r_{\mathrm{min}}$ (the transition region). In particular, $\sigma_{\mathrm{hh}}$ remains rather insensitive to $\eta^2$ on various $r$ scales. For light haloes, these recognizable effects of the asymmetry, particularly in the two-halo-group region, and the directionality of its effects on $v_{\mathrm{hh}}$ are consistent with the patterns observed in the context of $v_\mathrm{pp}$, as discussed in Section \ref{sc:particle-particle-pwv}.

For heavy haloes (solid lines in Figure \ref{fig:hh-pwv}), $r_{\mathrm{min}}\sim2\ \mathrm{Mpc}$, where the effects of halo merging are dominant. This is only slightly larger than the average virial radius (approximately $1\ \mathrm{Mpc}$) of these heavy haloes. Therefore, the result of multiple heavy haloes colliding and merging is observed. This behaviour is also reflected in $\sigma_{\mathrm{hh}}$, which has a similar peak on the same scale.

In addition, the $v_\mathrm{hh}$ for heavy haloes also show a transition point at approximately 8 Mpc, similar to light haloes. This suggests an intermediate zone between the one-halo-group and two-halo-group regions. When $r$ is larger than this transition range, $v_{\mathrm{hh}}$ and $\sigma_{\mathrm{hh}}$ have similar shapes as those of light haloes, but with larger magnitudes. In contrast, heavy haloes are less affected by neutrinos than light haloes, due to their greater gravitational attraction and resistance to smoothing.

Similar to the particle-particle case, we study the ratio of halo-halo mean pairwise peculiar velocity, $R^v_{\mathrm{hh}}$,
\begin{equation}
    R^v_{\rm{hh}}(M_\nu,\eta,r)
        \equiv
    \frac{v_{\mathrm{hh}}(M_\nu,\eta,r)}{v_{\mathrm{hh}}(0,0,r)}.
\end{equation}

We calculate the mean deviation of $R^v_{\mathrm{hh}}$ from 1, denoted as $\Delta R^v_{\mathrm{h}j}$,
\begin{equation}
    \Delta R^v_{\mathrm{h}j}(M_\nu,\eta) 
    \equiv 
    \left.\overline{ R^v_{\mathrm{hh}}(M_\nu,\eta,r) }\right\vert_{r\in I_j} - 1,
\end{equation}
where $j=1\ (2)$ represents the interval $I_1=[4,15]\ \mathrm{Mpc}$ ($I_2=[25,50]\ \mathrm{Mpc}$). For light haloes, we study both intervals, which allows us to investigate the effects of neutrinos in the transition region and the two-halo-group region. For heavy haloes, we focus on $I_2$, since the $I_1$ region has a high level of statistical noise, making it difficult to obtain reliable results with the current simulation resolution.

Similarly, we apply the two-variable linear regression model for both heavy and light haloes, which are expressed as $\Delta R^v_{\mathrm{h}j}=C^v_{m\mathrm{h}j} M_\nu'+C^v_{\eta\mathrm{h}j}\eta^2$. This approach is also applied to $\sigma_{\mathrm{hh}}$.

\begin{table*}
\large
\centering
  \begin{tabular}{ccccc}
    \hline
    \hline
    $M_{\mathrm{vir}}\ [M_\odot]$ & range [Mpc] & quantity 
    & $C^{v/\sigma}_{m\mathrm{h}j}$ (\%)& $C^{v/\sigma}_{\eta\mathrm{h}j}$ (\%)\\
    \hline
    \multirow{4}{*}{$[10^{11},10^{13}]$} 
    & \multirow{2}{*}{[4,15]} 
    & \multicolumn{1}{c}{$v_{\mathrm{hh}}$}      & \multicolumn{1}{c}{$-1.218\pm0.074$} & \multicolumn{1}{c}{$0.428\pm0.400$} \\
    &
    & \multicolumn{1}{c}{$\sigma_{\mathrm{hh}}$} & \multicolumn{1}{c}{$-1.471\pm0.046$} & \multicolumn{1}{c}{$1.705\pm0.253$} \\
    & \multirow{2}{*}{[25,50]} 
    & \multicolumn{1}{c}{$v_{\mathrm{hh}}$}      & \multicolumn{1}{c}{$-0.786\pm0.017$} & \multicolumn{1}{c}{$-3.389\pm0.089$} \\
    &
    & \multicolumn{1}{c}{$\sigma_{\mathrm{hh}}$} & \multicolumn{1}{c}{$-0.960\pm0.004$} & \multicolumn{1}{c}{$0.997\pm0.023$}  \\    
    \hline
    \multirow{2}{*}{$[10^{13},10^{15}]$} 
    & \multirow{2}{*}{[25,50]} 
    & \multicolumn{1}{c}{$v_{\mathrm{hh}}$}      & \multicolumn{1}{c}{$0.027\pm0.055$} & \multicolumn{1}{c}{$-4.216\pm0.288$}  \\
    &
    & \multicolumn{1}{c}{$\sigma_{\mathrm{hh}}$} & \multicolumn{1}{c}{$-0.703\pm0.025$} & \multicolumn{1}{c}{$-0.021\pm0.133$} \\
    \hline
  \end{tabular}
\caption{Linear regression fitting results of $\Delta R^v_{\mathrm{h}j}$ and $\Delta R^\sigma_{\mathrm{h}j}$ for different $M_{\mathrm{vir}}$.}
\label{tb:fitting-results}
\end{table*}

The results of our analysis of $\Delta R^v_{\mathrm{h}2}$ from light haloes and $\Delta R^\sigma_{\mathrm{h}2}$ from heavy haloes are shown in the lower left and right panels of Figure \ref{fig:pp-hh-v12-d12-fitting-results}, respectively. The fitting coefficients are provided in Table \ref{tb:fitting-results}.

When $M_\nu'$ and $\eta^2$ are of the order of $O(1)$, the mass has a greater impact on $v_\mathrm{hh}$ for light haloes in the transition region with $C_{m\mathrm{h}1}^v(\%)=-1.218\pm0.074$ and $C_{\eta\mathrm{h}1}^v(\%)=0.428\pm0.400$. However, in the two-halo-group region, the effect of $\eta^2$ becomes more prominent than that of $M_\nu$, with $C_{m\mathrm{h2}}^v(\%)=-0.786\pm0.017$ and $C_{\eta\mathrm{h}2}^v(\%)=-3.389\pm0.089$. For the light halo $\sigma_\mathrm{hh}$, the effects of $M_\nu$ and $\eta^2$ are comparable, but they act in opposite directions, with $C_{m\mathrm{h1}}^\sigma(\%)=-1.471\pm0.046$, $C_{\eta\mathrm{h1}}^\sigma(\%)=1.705\pm0.253$ in the range of $[4,15]\ \mathrm{Mpc}$, and $C_{m\mathrm{h}2}^\sigma(\%)=-0.960\pm0.004$, $C_{\eta\mathrm{h}2}^\sigma(\%)=0.997\pm0.023$ in the interval of $[25,50]\ \mathrm{Mpc}$.

In contrast, $\sigma_\mathrm{hh}$ for heavy haloes are less affected by neutrino effects than their lighter counterparts in the two-halo-group region ($C_{m\mathrm{h}2}^\sigma(\%)=-0.703\pm0.025$, $C_{\eta\mathrm{h}2}^\sigma(\%)=-0.021\pm0.133$), and is not even sensitive to $\eta^2$. However, for $v_\mathrm{hh}$, $C_{m\mathrm{h}2}^v(\%)=0.027\pm0.055$, $C_{\eta\mathrm{h}2}^v(\%)=-4.216\pm0.288$, the $\eta^2$  become more prominent than mass effects. In particular, $v_\mathrm{hh}$ is not very sensitive to $M_\nu$.

Hence, ideally, we can independently measure $\eta^2$ by accurately determining the halo mean pairwise peculiar velocity in the two-halo-group region for heavy haloes. This, combined with the $v_\mathrm{hh}$ for light haloes in both the intermediate and two-halo regions, would allow us to gain a better understanding of the neutrino properties. This would also further validate the estimates of $\eta^2$ derived from observations of heavy haloes.

%%%%%%%%%%%%%%%%%%%%%%%%%%%%%%%%%%%%%%%%%%%%%%%%%%%%%%%%%%%%%%%%%%%%%%%%%%%%%%%%%%%%%
\section{Conclusions}\label{sc:summary}
%%%%%%%%%%%%%%%%%%%%%%%%%%%%%%%%%%%%%%%%%%%%%%%%%%%%%%%%%%%%%%%%%%%%%%%%%%%%%%%%%%%%%

In this study, we investigate and quantify the effects of neutrino mass $M_\nu$ and asymmetry $\eta^2$ on the particle-particle (halo-halo) mean pairwise peculiar velocity $v_\mathrm{pp}$ ($v_\mathrm{hh}$) and dispersion $\sigma_\mathrm{pp}$ ($\sigma_\mathrm{hh}$). All our results are summarized in Tables \ref{tb:fitting-results-pp} and \ref{tb:fitting-results}. These effects are due to the combination of refitted cosmological parameters, neutrino free-streaming effects, and the modified expansion history of the universe.

Given these results, the degeneracy of $M_\nu$ and $\eta^2$ found in the matter power spectrum \citep{cartonzeng2019effects} and the halo merger tree \citep{hiuwing2021} can be disentangled by the halo-halo mean pairwise peculiar velocity in different regimes (transition and two-halo-group) for both heavy and light haloes. Our key conclusions are summarized below.

\hspace*{\fill}

(i) For light haloes ($[10^{11},10^{13}]\ M_\odot$) in the transition range ($[4,15]\ \mathrm{Mpc}$), neutrino mass effects overwhelm asymmetry effects, while in the two-halo-group range ($[25,50]\ \mathrm{Mpc}$), neutrino asymmetry effects dominate. This would help to disentangle the two effects. 

(ii) Furthermore, heavy haloes ($[10^{13},10^{15}]\ M_\odot$) serve as a further reconfirmation in the two-halo-group range ($[25,50]\ \mathrm{Mpc}$), where the effects of the neutrino asymmetry dominate.

\hspace*{\fill}

Moreover, in our upcoming paper, we measure the halo-halo mean pairwise peculiar velocity using galaxy observational data \citep[Cosmicflows-4,][]{tully2023ApJ}. We target to measure/constrain $M_\nu$ and $\eta^2$.

Although we visually depict the one-halo region ($r\lesssim2\ \mathrm{Mpc}$) in our plots, we do not perform a quantitative analysis in this particular range due to the lack of resolution in our simulations. Further studies may benefit from neutrino-involved simulations with higher resolution, which could provide a more comprehensive understanding of the dynamics in this range.

%%%%%%%%%%%%%%%%%%%%%%%%%%%%%%%%%%%%%%%%%%%%%%%%%%%%%%%%%%%%%%%%%%%%%%%%%%%%%%%%%%%%%
\section*{Acknowledgments}
%%%%%%%%%%%%%%%%%%%%%%%%%%%%%%%%%%%%%%%%%%%%%%%%%%%%%%%%%%%%%%%%%%%%%%%%%%%%%%%%%%%%%

The authors thank useful discussion with Jianxiong Chen, Hiu Wing Wong, Zhichao Zeng, and the anonymous referee for helpful comments. The computational resources used for the simulations in this work were kindly provided by the Chinese University of Hong Kong Central Research Computing Cluster. Furthermore, this research is supported by grants from the Research Grants Council of the Hong Kong Special Administrative Region, China, under Project No.s AoE/P-404/18 and 14300223. SL acknowledges the supports by the NSFC grant (No. 11988101) and the K. C. Wong Education Foundation. All plots in this paper are generated by \texttt{Matplotlib} \citep{matplotlib2007}, and we also use \texttt{SciPy} \citep{scipy2020} and \texttt{NumPy} \citep{numpy2020}.

%%%%%%%%%%%%%%%%%%%%%%%%%%%%%%%%%%%%%%%%%%%%%%%%%%%%%%%%%%%%%%%%%%%%%%%%%%%%%%%%%%%%%
\section*{Data availability}
%%%%%%%%%%%%%%%%%%%%%%%%%%%%%%%%%%%%%%%%%%%%%%%%%%%%%%%%%%%%%%%%%%%%%%%%%%%%%%%%%%%%%
The simulation data used in this article will be shared upon a reasonable request to the corresponding author.

%%%%%%%%%%%%%%%%%%%% REFERENCES %%%%%%%%%%%%%%%%%%

%%%%%%%%%%%%%%%%%%%%%%%%%%%%%%%%%%%%%%%%%%%%%%%%%%

%%%%%%%%%%%%%%%%% APPENDICES %%%%%%%%%%%%%%%%%%%%%

\appendix

%%%%%%%%%%%%%%%%%%%%%%%%%%%%%%%%%%%%%%%%%%%%%%%%%%%%%%%%%%%%%%%%%%%%%%%%%%%%%%%%%%%%%
\section{Neutrino-involved simulation}\label{sc:simu_details}
%%%%%%%%%%%%%%%%%%%%%%%%%%%%%%%%%%%%%%%%%%%%%%%%%%%%%%%%%%%%%%%%%%%%%%%%%%%%%%%%%%%%%

This section offers a brief overview of grid-based neutrino-involved simulations based on previous studies \cite{ali2013efficient,cartonzeng2019effects,hiuwing2021}.

Due to the high thermal velocities of neutrinos and their inability to form structures smaller than their free streaming scale, the overdensities of neutrinos ($\delta_\nu(a)$) and cold collisionless matter ($\delta_{c+b}(a)$) evolve differently. As a result, the total overdensity $\delta_t(a)$ can be written as
\begin{equation}\label{eq:over-density}
    \delta_t(a)=[1-f_\nu(a)]\delta_{c+b}(a)+f_\nu(a)\delta_\nu(a),
\end{equation}
where $f_\nu(a)\equiv\Omega_\nu(a)/(\Omega_{c+b}(a)+\Omega_\nu(a))$ and $\Omega_i(a)$ is the fractional energy density of the $i^{th}$ component at the scale factor $a$.

In each simulation step, the overdensity $\delta_{c+b}$ is known. To calculate $\delta_t$ in Eq. (\ref{eq:over-density}), we need $\delta_\nu$, which is proportional to $\int F_\nu(t,\boldsymbol{r},\boldsymbol{p})d^3p$, where $F_\nu$ is the neutrino distribution function. Therefore, we make use of the Vlasov equation for $F_\nu$,
\begin{equation}\label{eq:vlasov_phys}
    \frac{dF_\nu}{dt} = 
    \frac{\partial F_\nu}{\partial t} 
    + \frac{\partial F_\nu}{\partial \boldsymbol{r}}\cdot\frac{d\boldsymbol{r}}{dt} 
    + \frac{\partial F_\nu}{\partial \boldsymbol{p}}\cdot\frac{d\boldsymbol{p}}{dt}
    = 0.
\end{equation}

For simplicity, we transform the physical coordinates $(t,\boldsymbol{r},\boldsymbol{v})$ into the comoving coordinates $(s,\boldsymbol{x},\boldsymbol{u})$ based on equations $ds=dt/a^2$, $\boldsymbol{x}=\boldsymbol{r}/a$, and $\boldsymbol{u}=a\boldsymbol{v}-Ha\boldsymbol{r}$. Therefore, Eq. (\ref{eq:vlasov_phys}) can be written as
\begin{equation}
    \frac{1}{a^2}\frac{\partial F_\nu}{\partial s} 
	+ \frac{\boldsymbol{u}}{a^2}\cdot\frac{\partial F_\nu}{\partial \boldsymbol{x}}
	- \ddot{a}a\boldsymbol{x}\cdot\frac{\partial F_\nu}{\partial\boldsymbol{u}}
	+ a\dot{\boldsymbol{v}}\frac{\partial F_\nu}{\partial \boldsymbol{u}} = 0.
\end{equation}

Using the Friedmann equations, Poisson equation, and the two following identities
\begin{equation}
	\begin{aligned}
		\frac{\ddot{a}}{a} &= -\frac{4\upi G}{3}\bar\rho_t, \\
		\dot{\boldsymbol{v}} &= 
		      -Ga^2\int
		      \left[
		          \rho_t(s,\boldsymbol{x'})-\bar\rho_t(s)
		      \right]
		      \frac{\boldsymbol{x}-\boldsymbol{x'}}{|\boldsymbol{x}-\boldsymbol{x'}|^3}d^3x', \\
		\frac{4\upi\boldsymbol{x}}{3} &= \int\frac{\boldsymbol{x}-\boldsymbol{x'}}{|\boldsymbol{x}-\boldsymbol{x'}|^3}d^3x',\\
		\rho_t(s,\boldsymbol{x})-\bar\rho_t(s) &= 
            \bar\rho_{c+b}(s)\delta_{c+b}(s,\boldsymbol{x})+\bar\rho_\nu(s)\delta_\nu(s,\boldsymbol{x}),
	\end{aligned}
\end{equation}
where $\rho_t$ ($\bar\rho_t$) is the total (mean) energy density encompassing CDM ($c$), baryons ($b$), and massive neutrinos ($\nu$), we can get
\begin{equation}\label{eq:vlasov_comoving}
    \begin{split}
	\frac{\partial F_\nu}{\partial s} 
	+ \boldsymbol{u}\cdot\frac{\partial F_\nu}{\partial \boldsymbol{x}} 
	- Ga^4\frac{\partial F_\nu}{\partial\boldsymbol{u}}\cdot
	\int 
        & \left[
	\bar\rho_{c+b}(s)\delta_{c+b}(s,\boldsymbol{x'}) 
        \right.\\
        & +\left.
        \bar\rho_\nu(s)\delta_\nu(s,\boldsymbol{x'})
	\right]
        \frac{\boldsymbol{x}-\boldsymbol{x'}}{|\boldsymbol{x}-\boldsymbol{x'}|^3}d^3x'=0.        
    \end{split}
\end{equation}

The neutrino distribution function $F_\nu$ can be expanded as
\begin{equation}
    F_\nu(s,\boldsymbol{x},\boldsymbol{u})=f_\nu^0(u) + f_\nu'(s,\boldsymbol{x},\boldsymbol{u}),
\end{equation}
where $f_\nu^0(u)$ is the unperturbed Fermi-Dirac distribution, while $f_\nu'(s,\boldsymbol{x},\boldsymbol{u})$ ($\ll f_\nu^0(u)$) represents the small perturbations in the neutrino distribution generated by gravitational potentials. Consequently, Eq. (\ref{eq:vlasov_comoving}) can be linearized as
\begin{equation}
    \begin{split}
	\frac{\partial f_\nu'}{\partial s} 
	+ \boldsymbol{u}\cdot\frac{\partial f_\nu'}{\partial \boldsymbol{x}} 
	- Ga^4\frac{\partial f_\nu^0}{\partial\boldsymbol{u}}\cdot
	\int
        & \left[
	\bar\rho_{c+b}(s)\delta_{c+b}(s,\boldsymbol{x'})
        \right.\\
        & +\left.
        \bar\rho_\nu(s)\delta_\nu(s,\boldsymbol{x'})
	\right]
	\frac{\boldsymbol{x}-\boldsymbol{x'}}{|\boldsymbol{x}-\boldsymbol{x'}|^3}d^3x'=0.
    \end{split}
\end{equation}
By applying Fourier transformation and integrating over $s$ from the simulation initial time $s_i$ to a subsequent time $s$, we get
\begin{equation}\label{eq:FT_valsov}
\small
    \begin{aligned}
	& \tilde f_\nu'(s,\boldsymbol{k},\boldsymbol{u})
	+ 4\upi G \frac{i\boldsymbol{k}}{k^2}\cdot\frac{\partial f_\nu^0(u)}{\partial\boldsymbol{u}} \\
        & \times \int_{s_i}^s ds' a^4(s') e^{-i\boldsymbol{k}\cdot\boldsymbol{u}(s-s')}
        \left[
	\bar\rho_{c+b}(s')\tilde\delta_{c+b}(s',\boldsymbol{k})+\bar\rho_\nu(s')\tilde\delta_\nu(s',\boldsymbol{k})
	\right] \\
	& = \tilde f_\nu'(s_i,\boldsymbol{k},\boldsymbol{u})e^{-i\boldsymbol{k}\cdot\boldsymbol{u}(s-s_i)},
    \end{aligned}
\end{equation}
where the symbols with `$\sim$' denote the corresponding Fourier transformed variables, and the initial $\tilde f_\nu'(s_i,\boldsymbol{k},\boldsymbol{u})$ can be expanded by Legendre polynomials
\begin{equation}
    \tilde f_\nu'(s_i,\boldsymbol{k},\boldsymbol{u})
	=
    \sum_{l=0}^{\infty} i^l \tilde f_\nu'^{(l)}(s_i,\boldsymbol{k},u)P_l(\hat k\cdot\hat u),
\end{equation}
where the coefficients $\tilde f_\nu'^{(l)}(s_i,\boldsymbol{k},u)$ can be approximated as \citep{ali2013efficient}
\begin{equation}\label{eq:perturbed_nu_dist}
    \begin{aligned}
	\tilde f_\nu'^{(0)}(s_i,\boldsymbol{k},u) &= f_\nu^0(u)\tilde{\delta}_\nu(s_i,\boldsymbol{k}),\\
	\tilde f_\nu'^{(1)}(s_i,\boldsymbol{k},u) 
        &= \frac{df_\nu^0(u)}{du}k^{-1}a_i\tilde\theta_\nu(s_i,\boldsymbol{k}) \\
	&= -\frac{df_\nu^0(u)}{du}k^{-1}a_i^2H(a_i)\tilde{\delta}_\nu(s_i,\boldsymbol{k}),\\
	\tilde f_\nu'^{(l)}(s_i,\boldsymbol{k},u) &= 0\ (l \geq 2),
    \end{aligned}
\end{equation}
where $\tilde\theta_\nu(s_i,\boldsymbol{k})$ is the neutrino velocity divergence in $k$ space at the initial time, and the ultimate equality of the second equation is obtained under the assumption that $\tilde\delta_\nu\sim a$. This assumption is valid at the initial time of the simulation, usually during the matter-dominated era when the linear theory is applicable.

Integrating over $du^3$ on both sides of Eq. (\ref{eq:FT_valsov}), we obtain
\begin{equation}\label{eq:vlasov_final}
\resizebox{1.03\columnwidth}{!} 
{$
    \begin{aligned}
	\tilde{\delta}_\nu(s,\boldsymbol{k})
	&=  4\upi G \Phi[k(s-s')] \\
	&\times\int_{s_i}^s ds' [a(s')]^4 (s-s')
        \left[
	\bar\rho_{c+b}(s')\tilde{\delta}_{c+b}(s',\boldsymbol{k})+\bar\rho_\nu(s')\tilde{\delta}_\nu(s',\boldsymbol{k})
	\right] \\
	&+ \tilde{\delta}_\nu(s_i,\boldsymbol{k}) \Phi[k(s-s_i)] \left[1+(s-s_i)a_i^2H(a_i)\right],
    \end{aligned}
$}
\end{equation}
where $\Phi(q)$ is defined as 
\begin{equation}
    \Phi(q) \equiv \frac{\int du u^2 f_\nu^0(u)j_0(uq)}{\int du u^2 f_\nu^{0}(u)},
\end{equation}
where $j_0$ is the zeroth-order spherical Bessel function.

Hence, based on Eq. (\ref{eq:vlasov_final}), we can determine the evolution of $\tilde\delta_\nu(s,\boldsymbol{k})$ after obtaining $\tilde\delta_{c+b}(s,\boldsymbol{k})$ at each simulation time step. Subsequently, we can calculate $\tilde\delta_t(s,\boldsymbol{k})$ using Eq. (\ref{eq:over-density}) and continue the simulation. This approach allows us to take into account the effect of neutrinos on CDM particles through the long-range force.

To speed up our simulations, we apply the following equations to calculate $\Phi(q)$ and the mean energy density of neutrinos $\bar\rho_\nu$. Inspired by Eq. (C2) in \cite{ali2013efficient}, we adopt the following representation for $\Phi(q)$,
\begin{equation}\label{eq:phiq}
    \Phi(q,\xi_i)=\Phi_0(q)R(q,\xi_i),
\end{equation}
where $\xi_i$ is the degeneracy parameter of the $i^{th}$ neutrino mass eigenstate, and
\begin{equation}
    \begin{aligned}
        \Phi_0(q)&=\frac{1+0.0168q^2+0.0407q^4}{1+2.1734q^2+1.6787q^{4.1811}+0.1467q^8}, \\
        R(q,\xi) &=\frac{1+C_1(\xi_i)q^2+C_2(\xi_i)q^4}{1+C_3(\xi_i)q^2+C_4(\xi_i)q^4},\\
        C_1(\xi_i) &= 0.0838 - 0.1335\xi_i^2 + 0.0165\xi_i^4, \\
        C_2(\xi_i) &= 0.0840 - 0.0182\xi_i^2 + 0.0008\xi_i^4, \\
        C_3(\xi_i) &= 0.0840 + 0.0174\xi_i^2 - 0.0010\xi_i^4, \\
        C_4(\xi_i) &= 0.0840 + 0.0356\xi_i^2 - 0.0015\xi_i^4.     
    \end{aligned}	
\end{equation}

The mean neutrino energy density $\bar\rho_\nu$ is calculated from\footnote{K. T. Lau, private communication (unpublished)}
\begin{equation}
    \begin{aligned}
        \bar\rho_\nu(M_\nu,\eta,a) &= a^{-3.5}\bar\rho_\nu(M_\nu,\eta,a=1) \\
        &\times
            \left[
            \frac{ \cosh{(D_1-D_2\ln{\frac{M_\nu}{T_{\nu,0}}a})} }
                 { \cosh{(D_1-D_2\ln{\frac{M_\nu}{T_{\nu,0}}})} }
            \right]^{1/2D_2},
    \end{aligned}
\end{equation}
where $\bar\rho_\nu(M_\nu,\eta,a=1)$ is the mean neutrino energy density nowadays, while the parameters $D_1(\eta)$ and $D_2(\eta)$ are
\begin{equation}
    \begin{aligned}
        D_1(\eta) &= 2.038 + 0.020\eta^2, \\
        D_2(\eta) &= 0.906 + 0.002\eta^2.		
    \end{aligned}
\end{equation}

Within the scope of our simulation settings, the relative accuracy of our fitting formula $\Phi(q,\xi)$ is found to be within 1\% for $q\leq2$ and 3\% for the entire positive $q$ range, with an absolute accuracy better than $0.002$. Similarly, for $\bar\rho(M_\nu,\eta,a)$, the relative accuracy is established to be better than 0.6\%.

The code used in this paper is developed based on \cite{cartonzeng2019effects}.The major difference is that we include one more term $\tilde f'_{\nu}{}^{(1)}$ in Eq. (\ref{eq:perturbed_nu_dist}) which is included in \cite{ali2013efficient}. The technical improvement is that we have optimized the calculations as described above.

%%%%%%%%%%%%%%%%%%%%%%%%%%%%%%%%%%%%%%%%%%%%%%%%%%%%%%%%%%%%%%%%%%%%%%%%%%%%%%%%%%%%%
\section{Convergence tests}\label{sc:convergence}
%%%%%%%%%%%%%%%%%%%%%%%%%%%%%%%%%%%%%%%%%%%%%%%%%%%%%%%%%%%%%%%%%%%%%%%%%%%%%%%%%%%%%

%%%%%%%%%%%%%%%%%%%%%%%%%%%%%%%%%%%%%%%%%%%%%%%%%%%%%%%%%%%%%%%%%%%%%%%%%%%%%%%%%%%%%
\subsection{The number of CDM particles}\label{sc:pp_pert_part}
%%%%%%%%%%%%%%%%%%%%%%%%%%%%%%%%%%%%%%%%%%%%%%%%%%%%%%%%%%%%%%%%%%%%%%%%%%%%%%%%%%%%%

Given the large number of CDM particles (1024$^3$) used in our simulations, it is not feasible to include all of them in the calculation of particle-particle velocity and dispersion. Therefore, we investigate how our calculations are affected by the number of CDM particles used. Here, we take A1 in Table \ref{tb:simu_table}, with $M_\nu=0.06\ \mathrm{eV}$ and $\eta=0$ in the simulation set $S_1$.

\begin{figure}
    \includegraphics[width=.95\columnwidth]{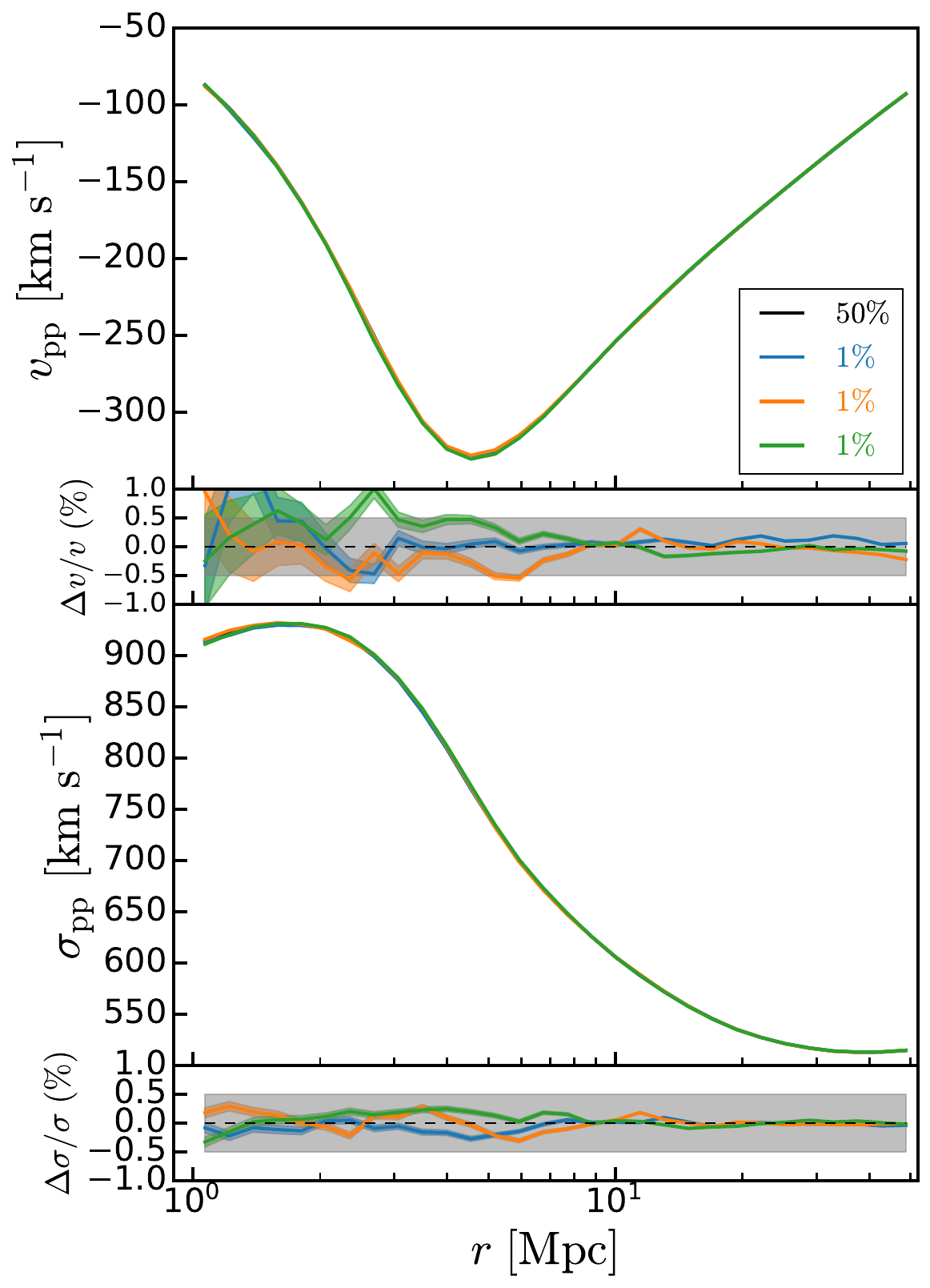}
    \caption{Particle-particle mean pairwise peculiar velocity ($v_{\mathrm{pp}}$, upper panels) and dispersion ($\sigma_{\mathrm{pp}}$, lower panels), respectively. The black line corresponds to a random selection of 50\% of the total CDM particles, while various coloured lines are different sets of randomly chosen 1\% particles. The lower subpanels, below $v_{\mathrm{pp}}$ and $\sigma_{\mathrm{pp}}$, show percentage fractional deviations with respect to the particle selection case 50\%, and the gray-shaded regions show deviations within $\pm$0.5\%.}
    \label{fig:part_used}
\end{figure}

Figure \ref{fig:part_used} shows four different scenarios for $v_\mathrm{pp}$ and $\sigma_\mathrm{pp}$. These scenarios involve using 50\% of the total CDM particles and three separate cases that involve randomly selecting 1\% of the CDM particles. We find that the differences in fractional percentages between the 1\% and 50\% cases are kept within 0.5\% in the region of interest ($r\gtrsim4\ \mathrm{Mpc}$).

Therefore, in this study, we use 1\% of the CDM particles for the $v_\mathrm{pp}$ and $\sigma_\mathrm{pp}$ calculations.

%%%%%%%%%%%%%%%%%%%%%%%%%%%%%%%%%%%%%%%%%%%%%%%%%%%%%%%%%%%%%%%%%%%%%%%%%%%%%%%%%%%%%
\subsection{Box size}\label{sc:pp_boxsize}
%%%%%%%%%%%%%%%%%%%%%%%%%%%%%%%%%%%%%%%%%%%%%%%%%%%%%%%%%%%%%%%%%%%%%%%%%%%%%%%%%%%%%

In this subsection, we select 1\% CDM particles for $v_\mathrm{pp}$ and $\sigma_\mathrm{pp}$ calculations, as discussed in Appendix \ref{sc:pp_pert_part}, to investigate how the effects of neutrinos are influenced by the box size.

\begin{figure}
    \includegraphics[width=.95\columnwidth]{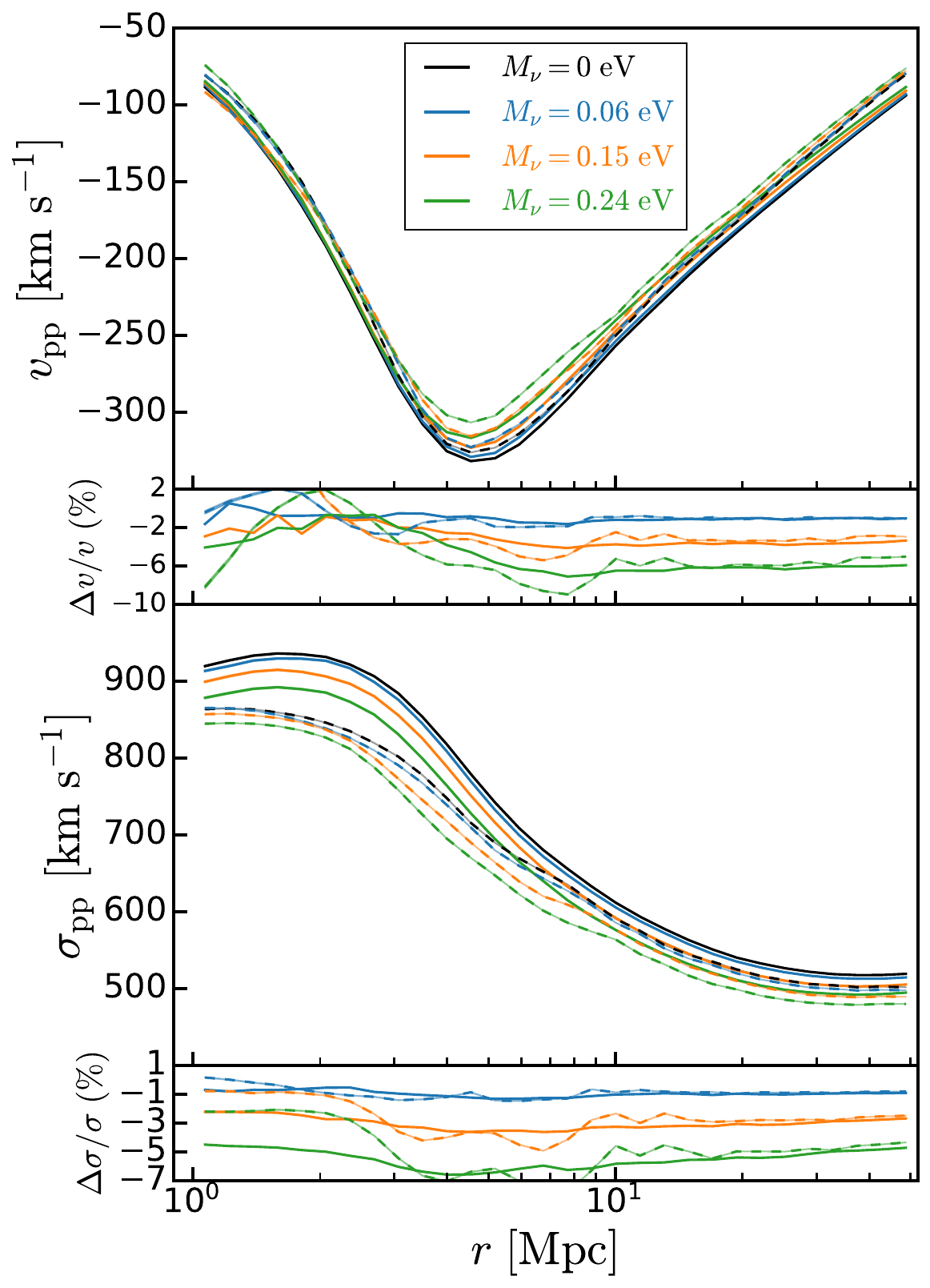}
    \caption{Particle-particle mean pairwise peculiar velocity ($v_{\mathrm{pp}}$, upper panels) and dispersion ($\sigma_{\mathrm{pp}}$, lower panels), respectively. Solid and dashed lines are from simulation sets $S_1$ and $S_2$, respectively. The various coloured lines correspond to $\eta^2=0$ and different $M_\nu$ (A1, A4, and A7 in Table \ref{tb:simu_table}). The lower subpanels show the percentage fractional differences from the fiducial massless case (A0).}
    \label{fig:pp_boxsize}
\end{figure}

The analysis, as shown in Figure \ref{fig:pp_boxsize}, keeps $\eta^2=0$ while changing $M_\nu$. Comparing the results from the simulation sets $S_1$ (solid lines) and $S_2$ (dashed lines) shows that although the magnitudes of $v_\mathrm{pp}$ and $\sigma_\mathrm{pp}$ are visibly different, the effects of neutrinos remain consistent beyond 4 Mpc, with only minor variations. These variations are mainly due to cosmic variance, with a slight contribution from the 1\% CDM particle subsample used. This result is in line with our expectation, since $v_\mathrm{pp}$ and $\sigma_\mathrm{pp}$ are statistical properties of CDM, and should not be too sensitive to the size of the simulation box. Similar results are seen when fixing $M_\nu$ while varying $\eta^2$.

Consequently, we choose the simulation set $S_1$ ($L_\mathrm{box}=1000\ h^{-1}\mathrm{Mpc}$) in our particle-related analysis.

%%%%%%%%%%%%%%%%%%%%%%%%%%%%%%%%%%%%%%%%%%%%%%%%%%%%%%%%%%%%%%%%%%%%%%%%%%%%%%%%%%%%%
\subsection{Cosmic variance}\label{sc:nu_cosmic_variance}
%%%%%%%%%%%%%%%%%%%%%%%%%%%%%%%%%%%%%%%%%%%%%%%%%%%%%%%%%%%%%%%%%%%%%%%%%%%%%%%%%%%%%

In this paper, we use a random seed value of 23456 for our simulations. Additionally, we conduct a set of neutrino-involved simulations with a different random seed, 12345, to assess the stability of neutrino effects. As described in Appendices \ref{sc:pp_pert_part} and \ref{sc:pp_boxsize}, only 1\% of the CDM particles in the simulation set $S_1$ are used to calculate $v_{\mathrm{pp}}$ and $\sigma_{\mathrm{pp}}$.

\begin{figure*}
    \includegraphics[width=.95\textwidth]{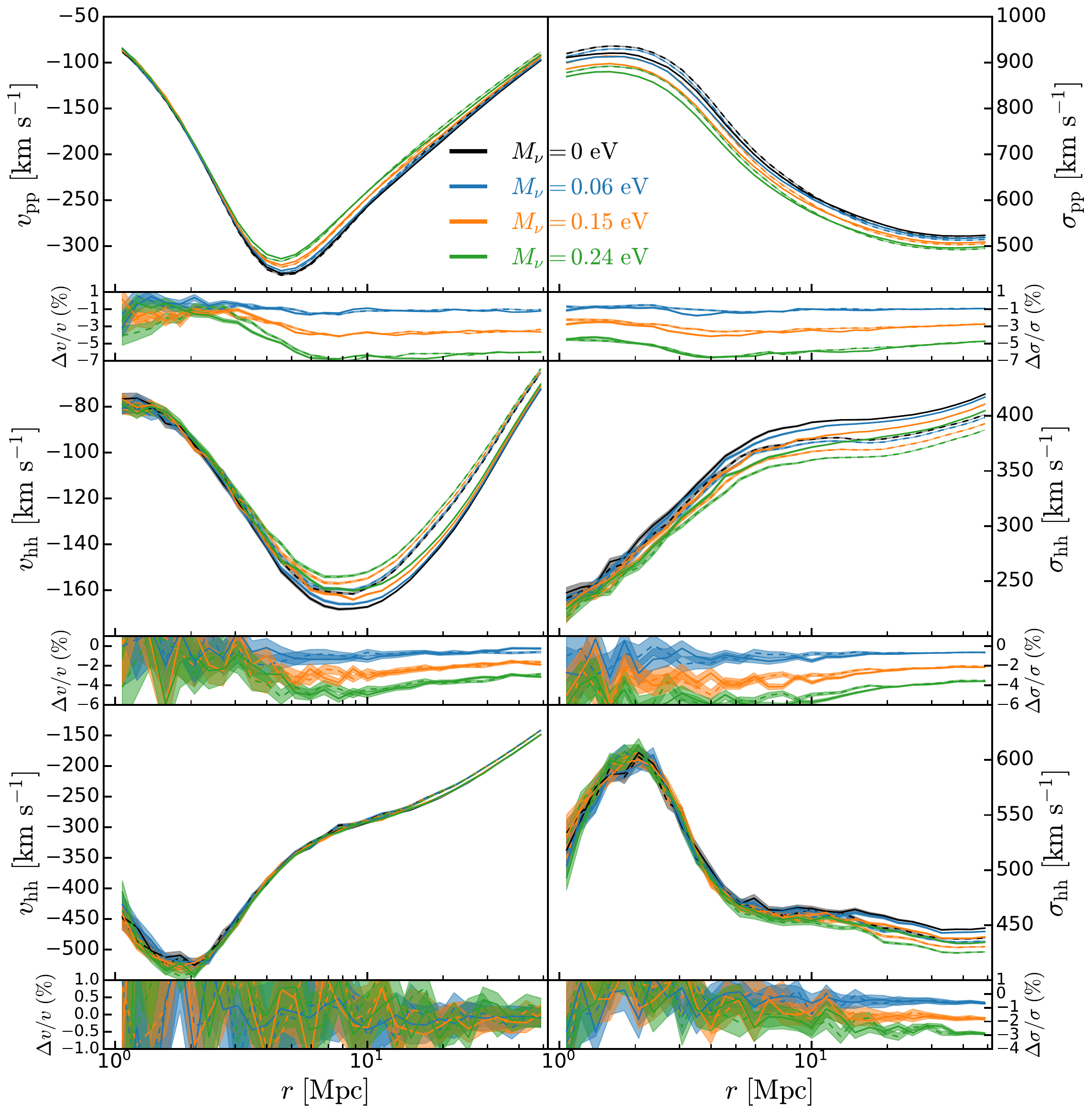}
    \caption{Particle-particle mean pairwise peculiar velocity ($v_{\mathrm{pp}}$, upper left panel) and dispersion ($\sigma_{\mathrm{pp}}$, upper right panel) computed from the set $S_1$, respectively, with percentage fractional deviations ($\Delta v/v$ and $\Delta \sigma /\sigma$) relative to the reference massless case (A0 in Table \ref{tb:simu_table}) presented in the subpanels below. The middle panels show the halo-halo mean pairwise peculiar velocity ($v_\mathrm{hh}$) and the corresponding dispersion ($\sigma_\mathrm{hh}$) calculated from light haloes ($[10^{11},10^{13}]M_\odot$), while the lower panels present $v_\mathrm{hh}$ and $\sigma_\mathrm{hh}$ for heavy haloes ($[10^{13},10^{15}]M_\odot$). Different random seeds are distinguished by solid and dashed lines. The colours in each panel represent different $M_\nu$ with $\eta^2=0$.}
    \label{fig:nu_cosmic_variance}
\end{figure*}

Figure \ref{fig:nu_cosmic_variance} shows the results, with solid and dashed lines representing the random seeds 23456 and 12345, respectively. Comparing lines of the same colour, we find that the absolute values of $v_{\mathrm{pp/hh}}$, $\sigma_{\mathrm{pp/hh}}$ may vary significantly, but the percentage fractional deviations ($\Delta v/v$, $\Delta\sigma/\sigma$) with respect to their respective fiducial massless cases (A0 in Table \ref{tb:simu_table}) remain stable between different random seeds within the range we are interested in ($r\gtrsim4\ \mathrm{Mpc}$). Additionally, this stability is also observed when fixing $M_\nu$ while varying $\eta^2$.

Consequently, in this paper, we will use seed 23456.

%%%%%%%%%%%%%%%%%%%%%%%%%%%%%%%%%%%%%%%%%%%%%%%%%%

\bsp	% typesetting comment
\label{lastpage}
\end{document}